\documentclass[fleqn,usenatbib]{mnras}

\usepackage{newtxtext,newtxmath}
\usepackage[normalem]{ulem}
\usepackage{listings}

\usepackage[T1]{fontenc}

\DeclareRobustCommand{\VAN}[3]{#2}
\let\VANthebibliography\thebibliography
\def\thebibliography{\DeclareRobustCommand{\VAN}[3]{##3}\VANthebibliography}

\usepackage{graphicx}	
\usepackage{amsmath}	
\usepackage{comment}
\usepackage{color}

\newcommand{\St}{\mathrm{St}}
\newcommand{\cs}{c_\mathrm{s}}
\newcommand{\ceff}{\tilde{c}_\mathrm{s}}
\newcommand{\Heff}{\tilde{H}}

\newcommand{\Vg}{V_\mathrm{gas}}
\newcommand{\Hp}{H_\mathrm{dust}}
\newcommand{\Omegav}{\Omega_\mathrm{vortex}}

\title[Planets and planetesimals at cosmic dawn]{Planets and planetesimals at cosmic dawn: Vortices as planetary nurseries} 

\author[L.E.J. Eriksson et al.]{
Linn E.J. Eriksson,$^{1,2}$\thanks{E-mail: leriksson@amnh.org}
Shyam Menon,$^{3,4}$
Daniel Carrera,$^{5}$
Wladimir Lyra,$^{5}$
Blakesley Burkhart$^{3,4}$
\\
$^{1}$ Department of Astrophysics, American Museum of Natural History, 200 Central Park West, New York, NY 10024, USA \\
$^{2}$ Institute for Advanced Computational Sciences, Stony Brook University, Stony Brook, NY, 11794-5250, USA \\
$^{3}$ Center for Computational Astrophysics, Flatiron Institute, 162 5th Avenue, New York, NY 10010, USA\\
$^{4}$ Department of Physics and Astronomy, Rutgers University, 136 Frelinghuysen Road, Piscataway, NJ 08854, USA\\
$^{5}$ New Mexico State University, Department of Astronomy, PO Box 30001 MSC 4500, Las Cruces, NM 88001, USA 
}

\date{Accepted XXX. Received YYY; in original form ZZZ}

\pubyear{2023}

\begin{document}
\label{firstpage}
\pagerange{\pageref{firstpage}--\pageref{lastpage}}
\maketitle

\begin{abstract}
Low-mass, metal-enriched stars were likely present as early as cosmic dawn. In this work, we investigate whether these stars could have hosted planets in their protoplanetary disks. If so, these would have been the first planets to form in the Universe, emerging in systems with metallicities much lower than solar.
In the core accretion model, planetesimals serve as the building blocks of planets, meaning that planetesimal formation is a prerequisite for planet formation. In a non-structured disk, planetesimal formation typically requires near-solar metallicities according to our current understanding. However, mechanisms that concentrate solid material can significantly lower this metallicity threshold.
Here, we explore whether vortices can facilitate the formation of the first planets and planetesimals at cosmic dawn. Vortices are prime sites for planetesimal formation due to their ability to efficiently trap and concentrate dust. We conduct simulations spanning a range of metallicities, and identify a metallicity threshold at $Z\gtrsim 0.04\, Z_{\odot}$ for planetesimal formation. If these planetesimals remain inside the vortex long enough to accrete the remaining trapped solids, Mercury-mass planets can form. The formation of Mars-mass planets or larger requires a metallicity of $Z\gtrsim 0.08\, Z_{\odot}$. These results assume a low level of disk turbulence, with higher turbulence levels leading to higher metallicity thresholds.

\end{abstract}




\section{Introduction}

The initial mass function (IMF) of stars forming in metal-poor environments (Population III/II) at cosmic dawn is a critical factor in shaping the early Universe's evolutionary landscape. In the absence of metals, primordial gas cools primarily through atomic and molecular hydrogen, leading to higher thermal pressures and Jeans masses that generally favor the formation of stars with masses ranging from tens to several hundred solar masses, i.e. a "top-heavy IMF" \citep[e.g.,][]{Klessen_2023}.  However, metal-free gas does not completely preclude the formation of low-mass stars, as dense pockets of gas in the discs of PopIII stars can undergo fragmentation to form low-mass stars -- an outcome realized in several numerical simulations \citep[e.g.,][]{Stacy_2014,Sharda_2020,Prole_2022,Sharda_2024}. Moreover, massive stars, with their short lifespans and energetic endpoints in supernovae, enrich the interstellar medium with heavy elements that provide efficient gas cooling channels for subsequent generations of star formation. These Population II stellar populations consist of a larger number of low-mass stars with its fraction increasing with metallicity \citep[e.g.,][]{Clark_2011,Chon_2021}, eventually transitioning to the bottom-heavy IMF characteristic of the local Universe at $Z \gtrsim 0.01 Z_{\odot}$ \citep{Sharda_2022}. 

The metal and dust enrichment in the early Universe is highly inhomogeneous in both space and time, governed by the interplay of star formation, feedback-driven outflows and inflowing metal-poor gas from the intergalactic medium. At cosmic dawn, the relatively young age of the Universe necessitates a stochastic, non-equilibrium approach to metal enrichment. Nevertheless, JWST has revealed that a mass-metallicity relation is already in place at $z \sim 5-10$, with metallicities as high as $0.3 Z_\odot$ in relatively more massive galaxies \citep[e.g.,][]{Curti_2024,Chakraborty_2024}. While this reflects gradual galactic-scale metal enrichment of the ISM, the distribution of metals is not uniform and localized regions may be present that have higher metallicities -- particularly within the dense gas of protogalactic disks or star-forming clumps. For instance, \citet{Whalen2008} show that supernova ejecta from even a single metal-free massive star can locally enrich dense gas to metallicities of $Z \sim 10^{-4}\,Z_{\odot}$, some of which show signatures of being gravitationally unstable, suggesting that localized metal-enriched star formation may persist out to even higher redshifts. The likely presence of several low-mass, metal-enriched stars at cosmic dawn offers the tantalizing possibility that they may form planets in their protoplanetary disks. 

In the first step of the planet formation process, $\sim \mu \textrm{m}$-sized dust and ice grains embedded in the gas collide and stick together, forming $\sim \textrm{mm}$-sized pebbles. Collisional growth beyond pebble size is impeded by several growth barriers. Collision experiments show that collisional velocities above $\sim 1-10\, \textrm{ms}^{-1}$ result in fragmentation rather than sticking \citep{BlumWurm2008, GundlachBlum2015, MusiolikWurm2019, Musiolik2021}, which is known as the fragmentation barrier. A second barrier, the radial drift barrier, arises because gas in the disk is pressure-supported and orbits at a sub-Keplerian velocity \citep{Weidenschilling1977}. In contrast, solid grains, which are not pressure-supported, would orbit at the Keplerian velocity (in the absence of gas drag). As grains grow and gradually decouple from the gas, they experience a headwind from the slower-moving gas, causing them to lose angular momentum and radially drift toward the star. The drift velocity depends on the degree of dust-gas coupling, governed by the Stokes number $\St$, with the highest drift velocities occurring for grains with $\St=1$. Other growth barriers include the bouncing barrier (e.g. \citealt{BlumMunch1993, Zsom2010, DominikDullemond2024}) and the charge barrier \citep{Okuzumi2011}, though neither will be considered in this study. 

Because collisional growth beyond pebble size is hindered by the previously mentioned barriers, an alternative mechanism must facilitate the formation of larger bodies.
This mechanism is gravitational collapse, which leads to the formation of $\sim 1-1000\, \textrm{km}$-sized planetesimals. For gravitational collapse to occur, the solids density must reach the so-called Roche density, which is significantly higher than the typical solids density in the disk. One favored mechanism for concentrating dust is the streaming instability (SI, \citealt{YoudinGoodman2005, JohansenYoudin2007}), which arises due to the mutual drag between solids and gas in the disk and leads to the formation of dense filaments that can locally reach Roche density. Much work has gone into identifying the conditions under which the SI leads to planetesimal formation \citep{Carrera2015,Yang2017,LiYoudin2021,Lim2025}. Based on these criteria, the SI would not operate in low-metallicity disks such as the ones considered in this work unless a separate mechanism first acts to concentrate the dust.

Vortices in the disk could provide the necessary mechanism for concentrating dust. They are equilibrium solutions of the Navier-Stokes equations characterized by closed elliptic streamlines, and have been proposed as prime sites for planetesimal formation due to their ability to efficiently trap dust (e.g., \citealt{BargeSommeria1995,AdamsWatkins95,Tanga+96, KlahrHenning97}). Grains that enter the vortex drift towards the pressure maximum at the center, where they accumulate, resulting in greatly enhanced solid densities \citep{GodonLivio00,Chavanis00,Johansen+04,InabaBarge06,KlahrBodenheimer06,Lyra2008,Lyra2009,Lyra2009b,ChangOishi10,Meheut+12,Fu+14,Raettig+15,Crnkovic-Rubsamen+15,Surville+16,Lyra+18,Raettig+21}. Studies of dust evolution inside vortices further show that grains grow to larger sizes at the vortex center \citep{Li2020}. Whether the SI operates in the same fashion inside a vortex as in a smooth protoplanetary disk is not fully known; however, preliminary work by \citet{Magnan2024} confirms that the drift of grains towards the vortex center powers an instability that closely resembles the traditional SI. Therefore, in this work, we proceed with the assumption that the SI criteria remain the same inside a vortex. Additionally, local 3D simulations with selfgravity to follow the gravitational collapse of the solids, performed by \citep{Lyra2024}, show that vortices are indeed efficient planetesimal factories, capable of forming objects with masses as large as those of the Moon and Mars.

Vortices have been invoked to explain the presence of crescent-shaped features in (sub)millimeter observations of protoplanetary disks on numerous occasions (e.g. \citealt{VanDerMarel2013, Perez2014, Fuente2017,Dong+18, Baruteau+19,Casassus+19,  vanderMarel+21}), suggesting that they might be common. Vortices can form for multiple reasons, such as through the Rossby wave instability \citep{Lovelace1999,Li+00,Li+01} at planetary gaps \citep{Tagger01, devalBorro+06,devalBorro+07,Lyra2009,Hammer+17}, at regions with strong gradients in mass accretion rate \citep{VarniereTagger06, Lyra2008,Lyra2009b,LyraMacLow12,Regaly2012,Lyra+15,Faure+15,Miranda2017,RegalyVorobyov2017}, through
the subcritic baroclinic instability \citep{KlahrBodenheimer2003,LesurPapaloizou10,LyraKlahr11}, the convective overstability \citep{KlahrHubbard14,Lyra14,Latter16, TeedLatter21,LehmannLin24,LehmannLin25}, and also as a result of a suggested nonlinear resonant buoyant instability \citep{Marcus+15,Marcus+16,Barranco+18}. Tentative evidence also links vortices to the saturated state of the Vertical Shear Instability \citep{Nelson+13,Richard+16,LatterPapaloizou18,MangerKlahr18}. Taken all in all, these results suggest that vortices are the end state of most purely hydrodynamical disk instabilities \citep{LyraUmurhan19,Lesur+23}. In this study, we operate under the premise that vortices are frequently occurring and investigate the implications for planet formation in low-metallicity disks.    

If planetesimals form inside a vortex, they will continue to grow through the accretion of pebbles \citep{Lyra2008,Lyra2009,Lyra2009b,Cummins2022}. In our low-metallicity disk, significant growth via the accretion of other planetesimals is highly unlikely, since planetesimal formation is likely rare and restricted to specific locations. Gas accretion can also be neglected, as the solid content of the disk is insufficient to form cores massive enough to trigger substantial gas accretion. However, planetesimals that form near the vortex center and do not immediately leave experience a period of greatly enhanced pebble accretion, as the remaining trapped pebbles are drifting towards vortex center \citep{Lyra2008,Lyra2009,Lyra2009b,Cummins2022}. This mechanism could potentially allow for the fast growth of planetary-mass objects via pebble accretion, even though pebble accretion rates outside the vortex are too low for meaningful growth. 

The aim of this study is to determine whether the presence of vortices can facilitate the formation of the first planets and planetesimals in very low metallicity disks. To this end, we perform simulations with a range of metallicities and identify the lowest metallicity at which planet formation becomes possible. In the context of planet formation, the term \textit{metallicity} is often loosely interpreted as the ratio of solid-to-gas mass in the disk. In models of dust evolution and planetesimal formation, the most commonly used parameter is the global dust-to-gas surface density ratio, $Z$. We adopt $Z$ as our definition of metallicity throughout this study. 

In Sect. \ref{sec: numerical methods}, we describe our models for dust evolution, planetesimal formation, and pebble accretion, as well as the post-processing that is used to emulate the effect of vortices. These models are applied in Sect. \ref{sec: formation of 1st planetesimals} to study when and where the first planetesimals might have formed. In Sect. \ref{sec: formation of 1st planets}, we investigate whether any of these planetesimals could serve as seeds for further growth into planets, assuming that growth occurs via pebble accretion. We discuss the limitations of our model and compare our results with previous studies in Sect. \ref{sec: discussion}, and finally summarize our results in Sect. \ref{sec: conclusion}. 

\section{Numerical method}\label{sec: numerical methods}
We consider a viscously evolving protoplanetary disk and simulate the growth and transport of dust within it. The effect of vortices is modeled via post-processing, and we compare our results with SI criteria to determine when and where planetesimals form. We then simulate the further growth of these planetesimals via pebble accretion, accounting for the enhanced accretion rate inside vortices. The code used and developed for this work is open-source and available on github\footnote{https://github.com/astrolinn/firstPlanets}.

\subsection{Evolution of the gas and dust disk}
We use the DustPy code by \citet{StammlerBirnstiel2022} to simulate the evolution of gas and dust in a 1D viscous protoplanetary disk. The evolution of the gas surface density and the transport of dust via radial drift is solved via the viscous advection-diffusion equation. The growth of dust is calculated by solving the Smoluchowski (coagulation) equation, where collisions lead to either perfect sticking or fragmentation depending on the collisional velocity (bouncing is not included in this work). In this study, we perform simulations using fragmentation velocities of $v_{\rm frag}=1\, \textrm{m}\, \textrm{s}^{-1}$ and $10\, \textrm{m}\, \textrm{s}^{-1}$. We adopt a dust bulk mass density of $1\, \textrm{g}\, \textrm{cm}^{-3}$ and conduct a parameter study exploring varying metallicities and turbulence levels. The amount of turbulence, characterized by the parameter $\alpha_{\rm turb}$, influences the frequency and velocity of dust collisions, as well as the dust scale height.

The initial gas surface density profile is given by,
\begin{equation}
    \Sigma_{\rm gas} = \frac{\dot{M}_0}{3\pi\nu_{\rm out}(r/r_{\rm out})^{\gamma}}\exp[-(r/r_{\rm out})^{2-\gamma}],
\end{equation}
where $\dot{M}_0$ is the initial disk accretion rate, $r$ is the semimajor axis, $\nu_{\rm out}$ is the kinematic viscosity at the exponential cut-off radius $r_{\rm out}$, and $\gamma$ is the radial viscosity gradient \citep{Lynden-Bell+1974}. The viscosity is given by, 
\begin{equation}
    \nu = \alpha \Omega_{\rm K}H^2,
\end{equation}
where $\alpha$ is the viscosity parameter \citep{ShakuraSunyaev1973}, $\Omega_{\rm K}$ is the Keplerian angular velocity and $H=c_{\rm s}/\Omega_{\rm K}$ is the gas scale height. The sound speed is given by,
\begin{equation}
    c_{\rm s} = \sqrt{\frac{k_{\rm B}T}{\mu m_{\rm H}}},
\end{equation}
where $k_{\rm B}$ is the Bolzmann constant, $T$ is the mid-plane temperature, $\mu$ is the mean molecular weight and $m_{\rm H}$ is the mass of the hydrogen atom. The temperature profile is taken to be that of a passively irradiated disk with an irradiation angle of $0.05$: 

\begin{equation}
    T = \sqrt[4]{\frac{1}{2}\frac{0.05L_*}{4\pi r^2 \sigma_{\rm SB}}}
\end{equation}
\begin{equation}
    L_* = 4\pi R_*^2 \sigma_{\rm SB} T_*^4
\end{equation}

\noindent where $L_*$, $R_*$, and $T_*$ are the luminosity, radius, and effective temperature of the central star, respectively, and $\sigma_{\rm SB}$ is the Stefan-Boltzmann constant \citep{ChiangGoldreich1997}.

We choose a disk lifetime of $3\, \textrm{Myr}$, at which point planetary growth is terminated. We adopt a semimajor axis grid spanning $1-1000\, \textrm{au}$, with 150 logarithmically spaced grid cells in the range $1-200\, \textrm{au}$ and 16 linearly spaced grid cells in the range $200-1000\, \textrm{au}$. The large radial extent is adopted to mitigate numerical issues near the outer radial boundary. The following parameter values are adopted throughout this work: $r_{\rm out}=25\, \textrm{au}$, $\gamma=1$, $\alpha=5\times 10^{-3}$ and $\mu=2.34$. 

We motivate our initial conditions from the primordial star formation simulations of \citet{Sharda_2024} -- specifically of a low-mass star that formed through fragmentation within the disc of a massive protostar, and is ejected from the disk via many-body interactions, subsequently evolving quasi-independently as an isolated star-disk system. We adopt its stellar mass of $0.45\, \textrm{M}_{\odot}$ and its corresponding initial disk mass of $0.0071\, \textrm{M}_{\odot}$ in our main study, which corresponds to an initial disk accretion rate of $\dot{M}_0=1.32\times 10^{-8}\, \textrm{M}_{\odot}\textrm{yr}^{-1}$ (calculated using a disk ranging from $1-200\, \textrm{au}$). We adopt stellar radii and effective temperatures from \citet{Baraffe2015} ($R_*=2.411\, \textrm{R}_{\odot}$ and $T_*=3761\, \textrm{K}$ for a $0.45\, \textrm{M}_{\odot}$ star). In the Appendix, we also present results from simulations performed with a solar mass star and a disk-to-star mass ratio of $0.1$ and $0.01$.

\subsection{Concentration and growth of dust in vortices}
In this section, we describe how the DustPy data is post-processed to obtain what the dust density and $\St$ would have been in the presence of vortices. 
We consider two distinct scenarios for dust concentration and growth inside a vortex. Case VC for "vortex concentration" represents a scenario where dust is concentrated inside the vortex, but there is no additional dust growth compared to the unperturbed disk. Case VCG for "vortex concentration plus growth" represents a scenario where both concentration and growth are promoted inside the vortex, leading to a feedback loop due to the interdependence of these processes.

The peak dust density obtained at the center of a Kida vortex\footnote{A Kida vortex is a patch of constant vorticity, producing a velocity field that smoothly transitions into the Keplerian shear.} \citep{Kida81} was derived analytically by \citet{Lyra2013,Lyra2024} as:
\begin{equation}\label{eq: vortex-boost}
    \epsilon_{\rm max} = \frac{\max(\rho_{\rm dust})}{\rho_{\rm gas}} = Z \left( \frac{\St}{\alpha_{\rm t}} + 1 \right)^{3/2}.
\end{equation}
This is a drift-diffusion equilibrium model, in which $\rho_{\rm dust}$ and $\rho_{\rm gas}$ are the dust and gas midplane density, respectively, and $Z$ is the dust-to-gas surface density ratio of the unperturbed disk. The above expression is derived for a mono-disperse dust size distribution. Since DustPy outputs an entire size distribution, we need to choose a representative $\St$, which we pick to be the density-weighted average $\St$. 

In Case VC, we post-process the DustPy data using Eq. \ref{eq: vortex-boost} to obtain what the peak dust-to-gas midplane density ratio would be at each point on the time and semimajor axis grid if a vortex were to form there. In this model, we do not consider the time required to form the vortex or the time needed to reach drift-diffusion equilibrium, but assume that it occurs instantaneously. Since we assume that the $\St$ remain the same as in the unperturbed disk - i.e., no growth or destruction of dust inside the vortex - there is no dust coagulation timescale involved. 

In Case VCG, we relax the last assumption and account for both the concentration and growth of dust inside the vortex, while also considering the effects of mass loading. The full derivation and implementation of this case are provided in \citet{Carrera2025}; here, we present a summary. Studies of dust growth inside vortices and pressure bumps find that grains grow to significantly larger sizes compared to the background disk \citep{Bae2018,Li2020}. Due to the lack of fast radial drift, it is reasonable to assume that growth proceeds until the fragmentation limit. When accounting for mass loading, we find that the expression for $\St_{\rm frag}$ depends positively on the dust density. Similarly, the expression for the peak dust density has a positive dependence on $\St$ (see Eq. \ref{eq: vortex-boost}). As a result, when dust concentrates at the center of the vortex, grains can grow to larger sizes, leading to even higher concentrations, creating a positive feedback loop. We note that, while theoretically motivated, the described feedback loop has yet to be numerically tested.

The concentration-growth feedback loop cannot go on indefinitely. There are three ways that it can terminate:

\begin{itemize}
\item Planetesimal formation: Once the dust reaches the Roche density, the dust is converted into planetesimals which ultimately leave the vortex.

\item Grain size limit: While there is no strict upper limit on $\St$, pebbles with $\St \gg 1$ will escape the vortex. Furthermore, the scaling relation for dust collision speeds changes for $\St>1$ \citep{Ormel_2007}, rendering our expression for $\St_{\rm frag}$ invalid. Therefore, we adopt $\St=1$ as an upper limit in this study.

\item Stagnation: For sufficiently small $\St$ and sufficiently large $\alpha$, the system can converge into a steady state
\end{itemize}

\noindent
Which of these end states is reached depends on the fragmentation barrier, strength of turbulence, as well as the relative timescales of dust growth $t_\St$ and vortex concentration $t_Z$. \citet{Birnstiel_2012} derived the dust growth timescale in a protoplanetary disk as $t_\St = 1/(Z\Omega)$. Here we follow \citet{Carrera2025} and assume that, since a vortex is in many ways analogous to a small protoplanetary disk, the dust growth timescale in a vortex is $t_\St = 1/(Z\Omegav)$, where $\Omegav$ is the vortex angular frequency. In practice, this has no meaningful effect on the simulation results since the two frequencies are almost identical. Following \citet{Lyra2013} and \citet{Carrera2025} we adopt $\Omegav = 0.5\Omega$. Finally, \citet{Carrera2025} derived the vortex concentration timescale $t_Z = 1/(\St \Omegav)$. Grain growth and vortex trapping occur in tandem, each on its respective timescale. For example, if drift-diffusion occurs on a shorter timescale, dust grains still grow toward $\St_{\rm frag}$, but not as quickly as the concentration moves toward drift-diffusion equilibrium. In this scenario, $\St_{\rm frag}$ is essentially a moving target for dust growth. We follow \citet{Carrera2025} and set the iteration timestep to $dt = \min(t_\St, t_Z)$, allowing the two processes to evolve gradually within a changing environment.

We assume that the vortex has a finite lifetime, which sets an upper limit on the time available for growth and concentration to proceed. We adopt a vortex lifetime of 1000 orbital periods in our main study. This is motivated by the re-generation of the m=1 mode of vortices from Rossby-Wave-Instability cycles found in \citet{Regaly2012}, which was found to be of the order $\sim$1000 orbits. We discuss how shorter or longer vortex lifetimes affect our results in Section \ref{subsec: lifetime vortex}. To keep the model simple and conservative, we assume that the vortex’s total dust budget remains constant after formation; in other words, we do not account for pebbles drifting into the vortex post-formation. 

We perform the post-processing at every point on the time and semimajor axis grid. The results should be interpreted as follows: for a vortex formed at $(r=r_j, t=t_i)$, the model outcomes $\St(j,i)$ and $\epsilon_{\rm max}(j,i)$ are the maximum Stokes number and dust-to-gas midplane density ratio that would be obtained at the vortex center during its 1000-orbit lifetime. Additional information regarding the derivation and implementation of case VCG is shown in Sect. \ref{sec:appendix:dust_growth}. 

Lastly, the total mass of dust trapped within a Kida vortex at any given time was given by \citet{Lyra2013} as:
\begin{equation}\label{eq: trapped dust}
    M_{\rm trapped} = (2\pi)^{3/2} \epsilon \rho_{\rm gas} \chi H H_g^2,
\end{equation}
where $\epsilon$ is the dust-to-gas density ratio of the background disk, $\chi=4$ is the vortex aspect ratio, $H_g=H/f(\chi)$ and $f(\chi)$ is a scale factor (see Eq. 35 of \citet{Lyra2013}). 

\subsection{Planetesimal formation}
To determine whether planetesimals are expected to form within the vortices, we compare the $(\St,\epsilon_{\rm max})$ from our two cases against the Roche density and the SI criterion. The Roche density is:
\begin{equation}
    \rho_{\rm R} = \frac{9\Omega_{\rm K}^2}{4\pi G}.
\end{equation}
The conditions under which the SI leads to planetesimal formation remains an active area of research. Perhaps the most reliable SI criterion comes from \citet{Lim2024}, who conducted a large number of 3D simulations of the SI under external (i.e., forced) turbulence. This represents a significant improvement over earlier criteria, which were based on 2D simulations without external turbulence. We summarize their criterion here, which we refer to as SI24:
\begin{equation}
    \log(\epsilon_{\rm crit}) \approx 0.42 (\log \St)^2
    + 0.72 \log \St + 0.37
\end{equation}
Importantly, \citet{Lim2024} showed that expressing the criterion in terms of $\epsilon$ instead of $Z$ makes the criterion remain valid for a range of alpha values (they tested $10^{-4} \le \alpha_{\rm t} \le 10^{-3}$). The one limitation of this work is that, due to the cost of 3D simulations, they could only model a limited range of Stokes numbers $(0.01 \le \St \le 0.1)$, while grains in in the disk can be much smaller than this range. 

To address this limitation, we also consider the criterion by \citet{Lim2025} (henceforth referred to as SI25), which is currently the most comprehensive among the 2D criteria available in the literature:
\begin{equation}\label{eq: SIb}
    \log(Z_{\rm crit}) \approx 0.10 (\log \St)^2
    + 0.07 \log \St - 2.36.
\end{equation}
For our analysis, we convert this surface density ratio criterion into its midplane dust-to-gas density ratio equivalent, $\epsilon_{\rm crit}$ (the derivation of $\epsilon_{\rm crit}$ is provided in Sect. \ref{sec:appendix:dust_growth:si_criterion}). 

The size distribution of planetesimals and embryos formed within a vortex is not fully known, since numerical simulations converges with resolution at the high mass end, but not at the low mass end \citep{Lyra2024}. Therefore, we consider the size distribution obtained from pure SI simulations. The typical mass of a planetesimal formed via the SI is:
\begin{equation}
    M_{\rm pl} = 5\times 10^{-5}\textrm{M}_{\oplus} \left(\frac{Z_{\rm fil}}{0.02}\right )^{1/2} \left(\frac{\zeta}{\pi^{-1}}\right )^{3/2} \left(\frac{h}{0.05}\right )^3 \left(\frac{M_*}{\textrm{M}_{\odot}}\right )   
\end{equation}
\citep{Liu2020,Lorek2022}. In the above equation $\zeta=4\pi G \rho_{\rm gas}\Omega_{\rm K}^{-2}$ is the self-gravity parameter, $h=H/r$ is the disk aspect ratio and $Z_{\rm fil}$ is the filament dust-to-gas ratio, here taken to be $0.1$.
The planetesimals most likely to continue growing into planets are those at the upper end of the size distribution. We follow \citet{Lyra2023}, who assumed that the most massive planetesimal is $\sim$10x more massive than the typical planetesimal mass. We refer to this object as an embryo, and further limit the embryo mass by the total mass of dust trapped within the vortex.

\subsection{Pebble accretion}
How much the embryo grows by pebble accretion is strongly influenced by the residence time of the embryo within the vortex (shown in Fig. \ref{fig: tcross}). Since we are considering very low dust-to-gas ratios in this work, the criteria for forming planetesimals can only be reached near the vortex center, where the dust density is high and the residence time is long. We adopt a nominal residence time of 10 orbital periods. We make the simplified assumption that each vortex forms only one embryo, and the remainder of the solid budget remains in dust and pebbles (see Sect. \ref{subsubsec: multiple embryos} for a discussion of this assumption).

Pebbles within the vortex drift towards the pressure maximum at the vortex center, resulting in a rapid accretion rate for embryos located near the center. \citet{Cummins2022} provide an expression for this "vortex-assisted" pebble accretion rate, which is independent of the $\St$ and the embryo mass. When implementing this accretion rate, we found that the embryo quickly accreted all of the trapped pebbles within the vortex. Therefore, we choose to simplify the accretion prescription by assuming that the embryo accretes all of the trapped pebbles during the first orbital period after formation. Later, the maximum possible accretion rate is the rate at which pebbles drift into the vortex. We make the assumption that the embryo continues accreting at this maximum rate, for the remainder of the embryo's residence time within the vortex. 

After the embryo leaves the vortex, we assume that pebble accretion proceeds in the standard fashion, described in \citet{Lyra2023}. The exact mono-disperse pebble accretion rate is:
\begin{equation}
    \dot{M}_{\rm pebb} = \pi R_{\rm acc}^2 \rho_{\rm dust} \delta v e^{-\xi}[I_0(\xi)+I_1(\xi)],
\end{equation}
where
\begin{equation}
    \xi \equiv \left(\frac{R_{\rm acc}}{2H_{\rm dust}}\right)^2.
\end{equation}
In the above expressions $R_{\rm acc}$ is the accretion radius, $\delta v$ is the approach velocity, $I_0(\xi)$ and $I_1(\xi)$ are the modified Bessel functions of the first kind, and $H_{\rm dust}$ is the dust scale-height (see \citet{Lyra2023} for the full set of equations).
To obtain the poly-disperse accretion rate, we calculate the mono-disperse accretion rate for all dust sizes and sum up the results. We note that since the accretion regime (Hill/Bondi) depends on grain size, and we consider an entire size distribution, there can be both Bondi and Hill accretion ongoing simultaneously. The standard 2D limit (Eq. 30 of \citet{Lyra2023}) is used when $\xi>10$ to prevent numerical issues. 

The planetary masses are updated using a simple Euler method with a linear time-step of $500\, \textrm{yr}$, and standard pebble accretion proceeds from the time the embryo leaves the vortex until the disk disperses at $3\, \textrm{Myr}$. In our model for growth via pebble accretion, we assume that the time between vortex formation and embryo formation is negligible, so the embryo’s formation time is taken to be the same as that of the vortex. In case VCG, we assume a vortex lifetime of 1000 orbital periods but do not track when during that period the SI criteria are first met. Because of this, we maintain our assumption for the embryo formation time in case VCG as well. However, we note that this assumption may not hold at large semimajor axes if the SI criteria are only met late in the vortex’s lifetime. As shown in the results, most significant growth occurs through the accretion of trapped pebbles, which happens very quickly after embryo formation. Therefore, our assumption regarding the embryo formation time should not have a significant impact on the results.

\section{Formation of the first planetesimals}\label{sec: formation of 1st planetesimals}

\begin{figure}
    \centering
    \includegraphics[width=\columnwidth]{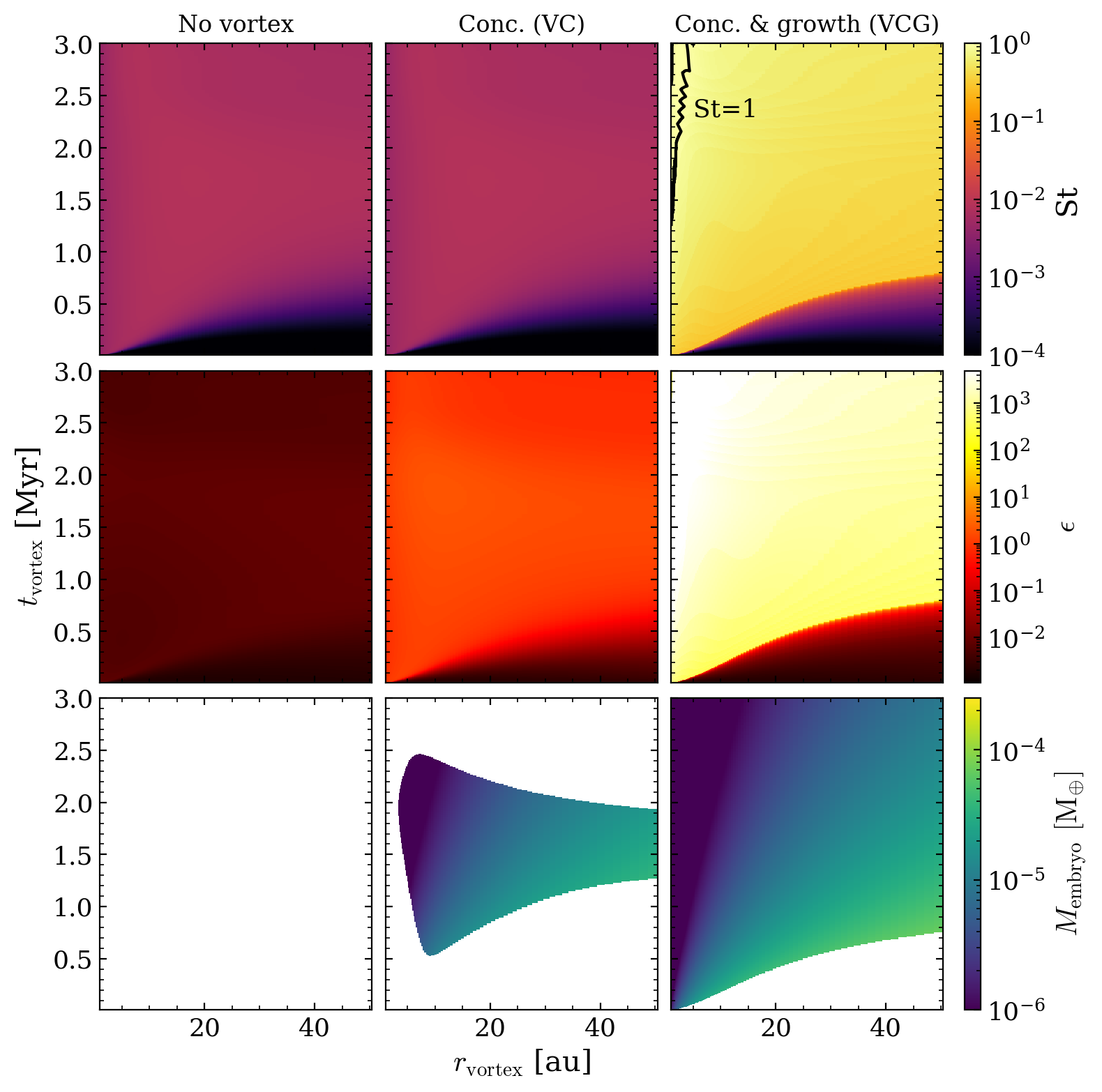}
    \caption{The plot shows the representative Stokes number (top row), midplane dust-to-gas density ratio $\epsilon$ ($\epsilon_{\rm max}$ for case VC/VCG, middle row), and planetesimal formation regions along with the resulting embryo masses (bottom row) for our simulation with $v_{\rm frag}=1\, \textrm{m}\, \textrm{s}^{-1}$, $Z=1.6\times 10^{-3}$, and $\alpha_{\rm T}=5\times 10^{-5}$. If there are no vortices (left column), $\epsilon$ is much too low to trigger planetesimal formation. The middle column shows the $\epsilon_{\rm max}$ and $\St$ that would be obtained at the center of a vortex formed at each individual space-time coordinate ($r_j, t_i$), assuming vortices act to concentrate the dust without affecting dust growth. In this scenario, $\epsilon_{\rm max}$ is significantly enhanced compared to the unperturbed disk, and vortex formation would result in planetesimal formation in large parts of the parameter space. If both concentration and growth are enhanced inside the vortex (right column), vortex formation would lead to planetesimal formation almost regardless of when are where the vortex forms.}
    \label{fig: disk_single}
\end{figure}

\begin{figure*}
    \centering
    \includegraphics[width=1\columnwidth, trim=0.2cm 0.2cm 2.8cm 0, clip]{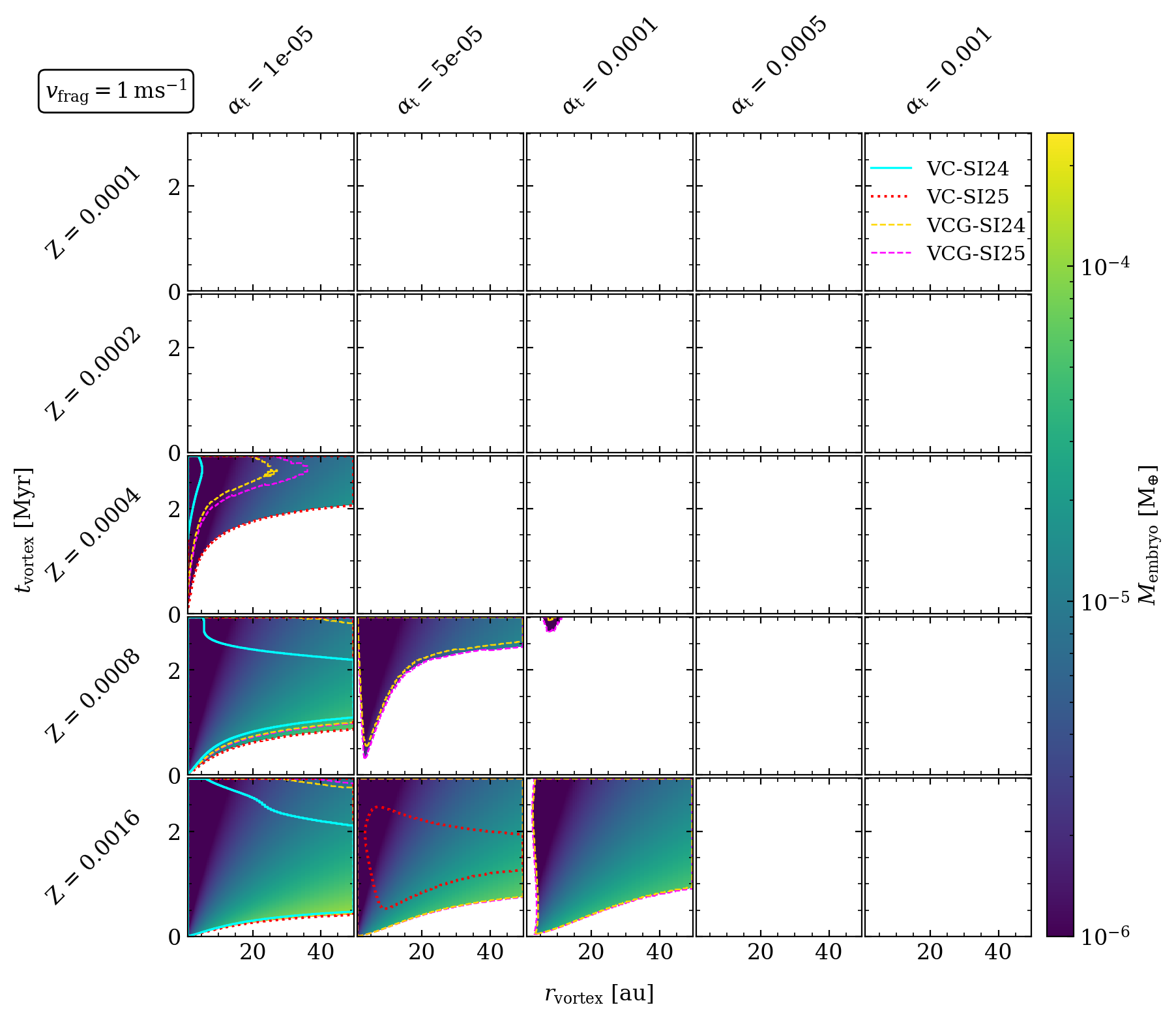}
    \includegraphics[width=1\columnwidth, trim=2.8cm 0.2cm 0.2cm 0, clip]{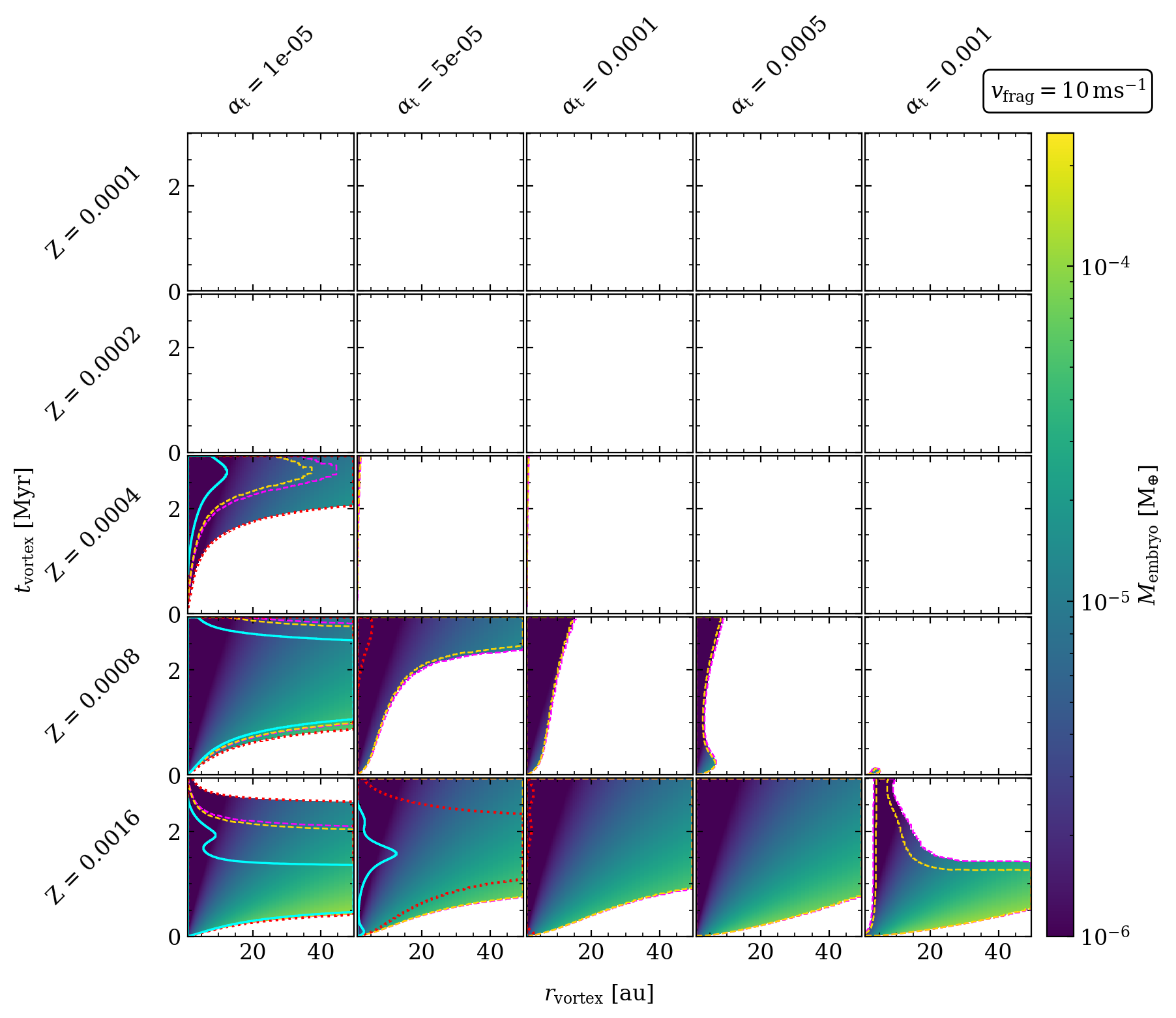}
    \caption{The plots show where on the space-time grid vortex formation would result in planetesimal formation for all simulations in the parameter study, along with the corresponding embryo masses. The colored lines outline the regions where planetesimals form in case VC and VCG, for both criteria SI24 and SI25.}
    \label{fig: membryo}
\end{figure*}

We perform a suite of DustPy simulations with varying initial $Z$, $\alpha_{\rm t}$, and $v_{\rm frag}$. The average metallicity of the Universe gradually increases over time; however, the spread in metallicity between different star-forming regions can vary significantly. Even at very high redshifts, there may be local regions that are enriched enough to trigger the planet formation process. In this work, we explore initial values of $Z$ in the range $10^{-4}-1.6\times 10^{-3}$, which, for an initial disk mass of $0.0071\, \textrm{M}_{\odot}$, corresponds to $0.24-3.8\, \textrm{M}_{\oplus}$ of solids. We consider $\alpha_{\rm t}$ in the range of $10^{-5}-10^{-3}$, which aligns with the values derived from protoplanetary disk observations \citep{Villenave+22,Pinte+23}.

The results from our simulation with $Z=1.6\times 10^{-3}$, $\alpha_{\rm t}=5\times 10^{-5}$, and $v_{\rm frag}=1\, \textrm{ms}^{-1}$ are shown in Fig. \ref{fig: disk_single}. The left column presents the evolution of the representative Stokes number (i.e., the density-weighted average $\St$) and the dust-to-gas midplane density ratio across the disk for the unperturbed case (i.e., the pure DustPy data). In the innermost disk, out to a few au, grains grow to the fragmentation limit within $\sim 10\, \textrm{kyr}$, resulting in $\St\sim 10^{-2}$. Further out, growth is limited by the radial drift barrier. The growth timescale increases with semimajor axis, and at $50\, \textrm{au}$, it takes $\sim 1\, \textrm{Myr}$ for grains to reach the radial drift limit. 
In this unperturbed scenario, the dust-to-gas midplane density ratio never exceeds $10^{-2}$, and no planetesimal formation occurs. 

The middle column of Fig. \ref{fig: disk_single} shows the $\St$ and peak dust-to-gas midplane density ratio $\epsilon_{\rm max}$ that would be obtained at the center of a vortex formed at each individual space-time coordinate ($r_j, t_i$), if vortices act to concentrate the dust without affecting dust growth (case VC). The bottom plot shows where in time and semimajor axis space vortex formation would result in planetesimal formation (here we use Eq.\ref{eq: SIb} for the SI criteria), along with the masses of the formed embryos. 
In case VC, $\epsilon_{\rm max}$ exceeds the SI criterion in large parts of the parameter space. Due to the low mass of the star and disk, the masses of the forming embryos are relatively small, on the order of $10^{-5}\, \textrm{M}_{\oplus}$. 

The right column of Fig. \ref{fig: disk_single} presents the results for case VCG, where both concentration and growth are enhanced inside the vortex, resulting in a growth-concentration feedback loop due to the interdependence of these processes. The presented $\St$ and $\epsilon_{\rm max}$ are the maximum Stokes number and dust-to-gas midplane density ratio that would be obtained at the center of a vortex during its 1000-orbit lifetime. In this scenario, vortex formation would result in planetesimal formation almost everywhere on the space-time grid, except at large semimajor axes when the disk is still young. This is because the dust densities and sizes are too small at the time of vortex formation, resulting in long growth and concentration timescales inside the vortex. The sharp transition between the region where the feedback loop is efficient and inefficient is highly dependent on the assumed vortex lifetime (see Fig. \ref{fig:appendix:embryo tvortex}). The black contour in the $\St$ plot outlines the region where $\St \geq 1$ is reached before the Roche density. 

Fig. \ref{fig: membryo} shows where on the space-time grid vortex formation would result in planetesimal formation for all simulations in the parameter study, along with the corresponding embryo masses.
In the scenario where dust sizes remain the same as in the unperturbed disk and concentration occurs instantaneously (case VC), vortices can trigger planetesimal formation for metallicities as low as $Z=0.0004$ if $\alpha_{\rm t}=10^{-5}$. For $\alpha_{\rm t}=5\times 10^{-5}$, a dust-to-gas ratio of $Z=0.0008$ or $Z=0.0016$ is required, depending on the assumed fragmentation velocity and SI criteria. At higher turbulence levels, merely concentrating dust inside vortices is no longer sufficient to trigger planetesimal formation. The exact space-time region where vortex formation would result in planetesimal formation depends on the assumed SI criteria. Criterion SI25 allows for planetesimal formation inside vortices in a wider region of parameter space than criterion SI24, which is expected, since criterion SI24 accounts for the effects of turbulence.

In the scenario where dust growth and concentration boost each other in a feedback cycle (case VCG), planetesimal formation inside vortices is also possible for $\alpha_{\rm t} > 5\times 10^{-5}$. When considering a fragmentation velocity of $1\, \textrm{ms}^{-1}$, planetesimal formation inside vortices is possible for $\alpha_{\rm t} \leq 10^{-4}$. Using a fragmentation velocity ten times larger allows for planetesimal formation even when $\alpha_{\rm t} > 10^{-4}$, provided $Z\gtrsim 0.0008$. Increasing the fragmentation velocity beyond $1\, \textrm{ms}^{-1}$ has little effect when $\alpha_{\rm t} < 10^{-4}$, since particle growth in this case is mostly limited by radial drift.

When considering $\alpha_{\rm t} = 10^{-5}$ and SI criterion SI25, case VCG results in narrower planetesimal formation regions than case VC. This is because case VCG accounts for the growth and concentration timescales inside the vortex, whereas case VC assumes that drift-diffusion equilibrium is reached instantaneously. Furthermore, the results of case VCG are highly dependent on the assumed vortex lifetime. A longer vortex lifetime - or equivalently, shorter growth and concentration timescales - would allow for additional dust growth and concentration. Fig. \ref{fig:appendix:embryo tvortex} shows how the region where vortex formation results in planetesimal formation changes when assuming a vortex lifetime that is 10 times longer. In this scenario, planetesimal formation becomes possible even for $Z=0.0002$.

Our combined results suggest that the presence of vortices could trigger planetesimal formation in disks with metallicities significantly lower than solar. We find a metallicity threshold of $Z\gtrsim 0.04\, Z_{\odot}$, which could extend to even lower metallicities in case VCG if the assumed vortex lifetime is significantly increased. Figure \ref{fig:appendix:membryo} shows that the obtained metallicity threshold also holds for a system with a solar-mass star and varying disk-to-star mass ratios. 

\section{Formation of the first planets}\label{sec: formation of 1st planets}

\begin{figure}
    \centering
    \includegraphics[width=0.9\columnwidth]{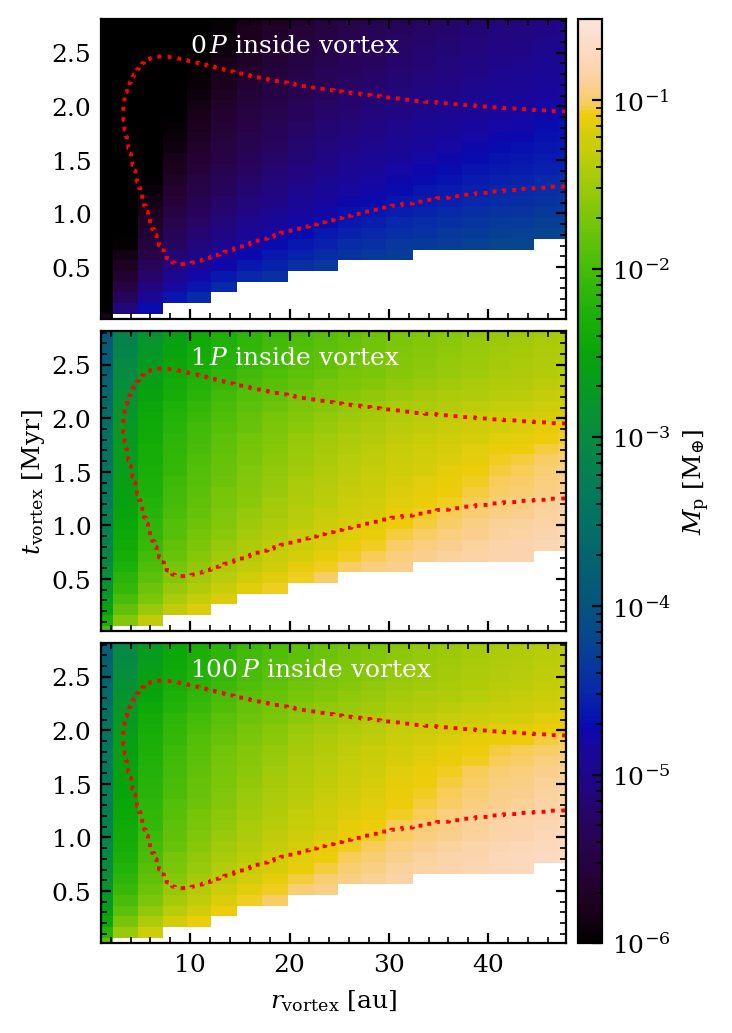}
    \caption{Final masses of planets forming inside a vortex as a function of the formation time and semimajor axis of the vortex, for the same simulation presented in Fig. \ref{fig: disk_single}. The three panels employ different assumptions for the embryos' residence time within the vortex. The results shown are for case VCG, with the red dotted line outlining the region where planetesimals form in case VC.}
    \label{fig: mp_single}
\end{figure}

In this section, we address whether the embryos that form are likely to act as seeds for planet formation, or if even higher metallicities are required for planetary objects to form. To save time and computational resources, we use a coarser space-time grid in this part of the study. For each point on the grid where vortex formation would result in planetesimal formation, we calculate the embryo mass (taken to be 10x the typical planetesimal mass) and simulate growth via pebble accretion onto this embryo. Fig. \ref{fig: mp_single} shows the resulting planetary masses at the end of the disk lifetime for the same simulation presented in Fig. \ref{fig: disk_single}. The different panels employ different assumptions for the embryos' residence time within the vortex. The results should be interpreted as follows: if a vortex were to form at $(r=r_j, t=t_i)$, the embryo that forms inside would grow to a mass $M_{\rm p}(j,i)$ by the end of the disk lifetime. 

If the embryo leaves the vortex immediately after formation (top plot), there is essentially no mass gain due to pebble accretion. The pebble accretion rate onto the embryo is very low, because of the small mass of the embryo and the low solid content of the disk. We find that this result holds true for all simulations in the parameter study. If the embryo remains inside the vortex just long enough to accrete all of the trapped dust (middle plot), the final planetary mass can reach $M_{\rm p}>0.1\, \textrm{M}_{\oplus}$. The majority of this mass was accreted inside the vortex, with little or no contribution from pebble accretion outside the vortex, except for in the innermost disk. However, even in the innermost disk, the contribution from standard pebble accretion is still less than the contribution from the accretion of trapped dust.

The bottom plot of Fig. \ref{fig: mp_single} shows the final planetary masses when the embryos reside inside the vortex for 100 orbital periods. In this scenario, the embryo accretes all of the trapped dust and experiences an additional 99 orbital periods of vortex-assisted pebble accretion. This results in a $\sim 15-30\%$ increase in the final planetary masses compared to when the residence time was one orbital period. These results show that the accretion of trapped dust is the most important contributor to planet formation inside vortices in low-metallicity disks, with subsequent pebble accretion playing a lesser role.

\begin{figure*}
    \centering
    \includegraphics[width=1\columnwidth, trim=0.2cm 0.2cm 2.8cm 0, clip]{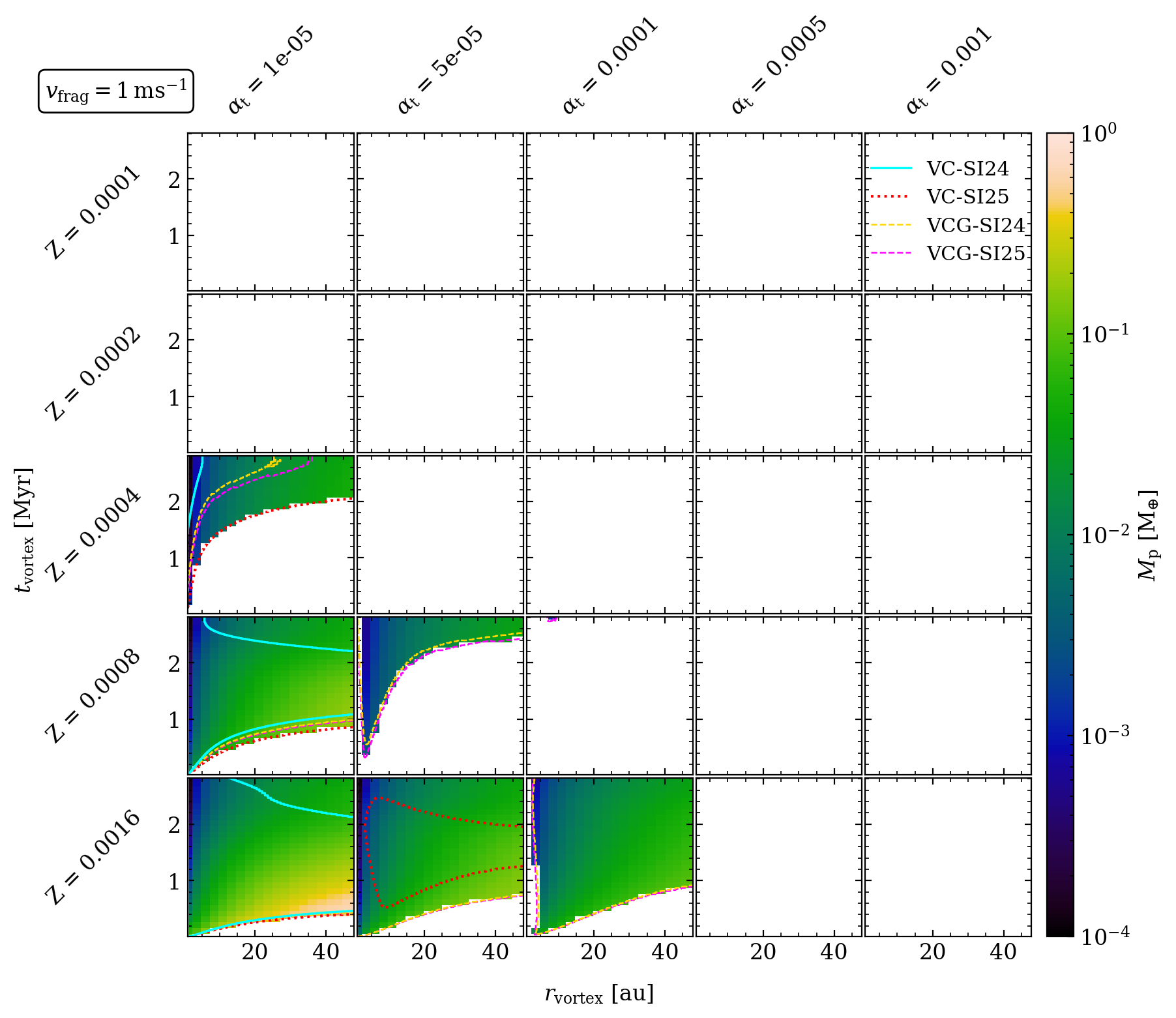}
    \includegraphics[width=1\columnwidth, trim=2.8cm 0.2cm 0.2cm 0, clip]{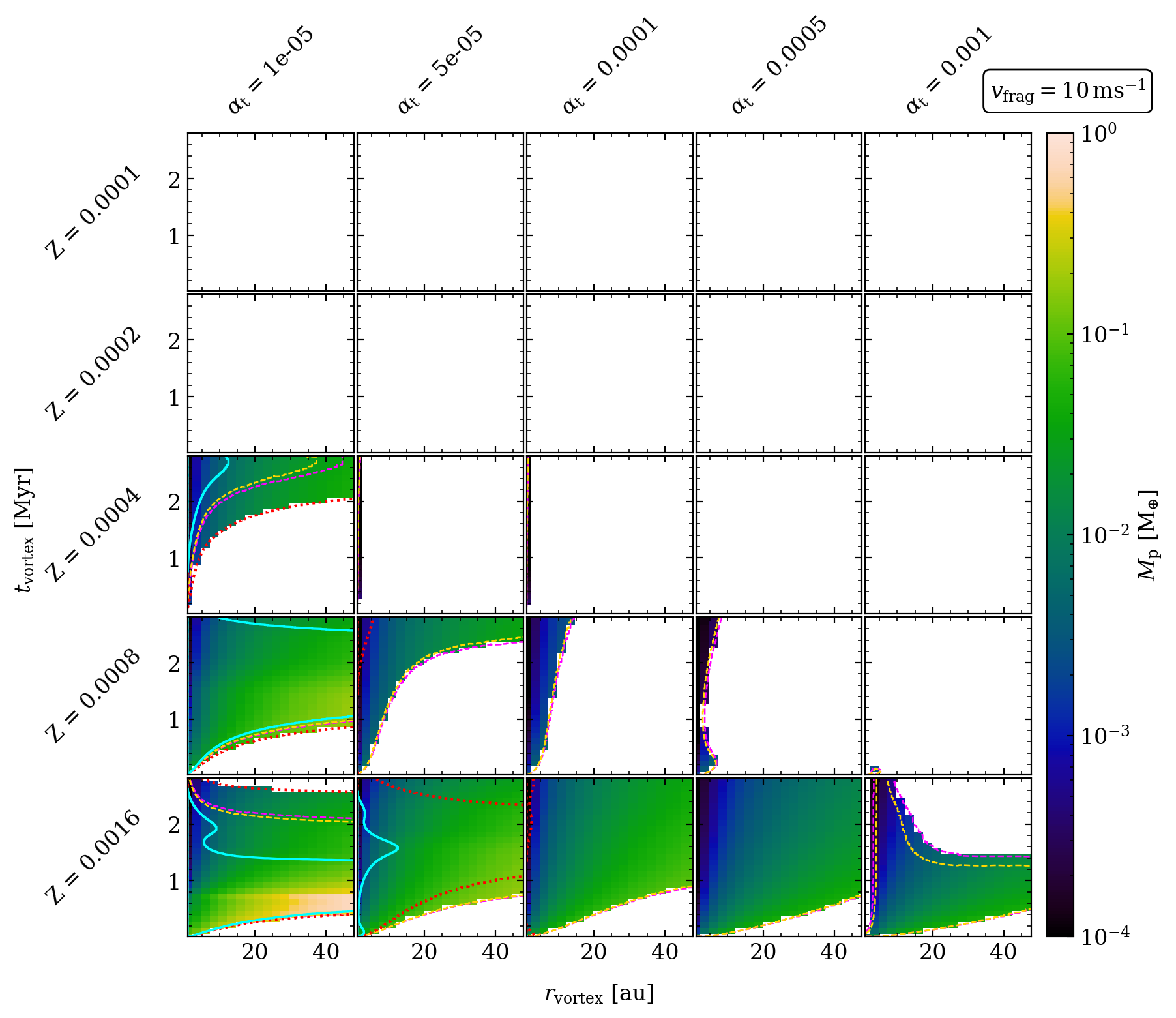}
    \caption{The plots show the final masses of planets forming inside a vortex as a function of the formation time and semimajor axis of the vortex, for all simulations in the parameter study. In these simulations it was assumed that the embryos reside ten orbital periods within the vortex, after which they continue to accrete at the normal accretion rate.}
    \label{fig: mp}
\end{figure*}

Fig. \ref{fig: mp} shows the planetary masses at the end of the disk lifetime for all simulations in the parameter study, assuming a residence time of ten orbital periods within the vortex. Generally, vortices that form in disks with higher $Z$ and lower $\alpha_{\rm t}$ produce planets with larger masses. The largest planets formed in our study have masses of $\sim 0.7\, \textrm{M}_{\oplus}$ and occur when $Z=1.6\times 10^{-3}$ and $\alpha_{\rm t}=10^{-5}$. When $Z=4\times 10^{-4}$ and $\alpha_{\rm t}=10^{-5}$, we obtain planets with masses just below that of Mercury. We note that these results are heavily dependent on the calculation of the trapped dust mass within the vortex (Eq. \ref{eq: trapped dust}) and the assumption that all trapped dust is accreted by the planet. 

\begin{figure*}
    \centering
    \includegraphics[width=0.9\textwidth, trim=0.2cm 0.2cm 0.3cm 0, clip]{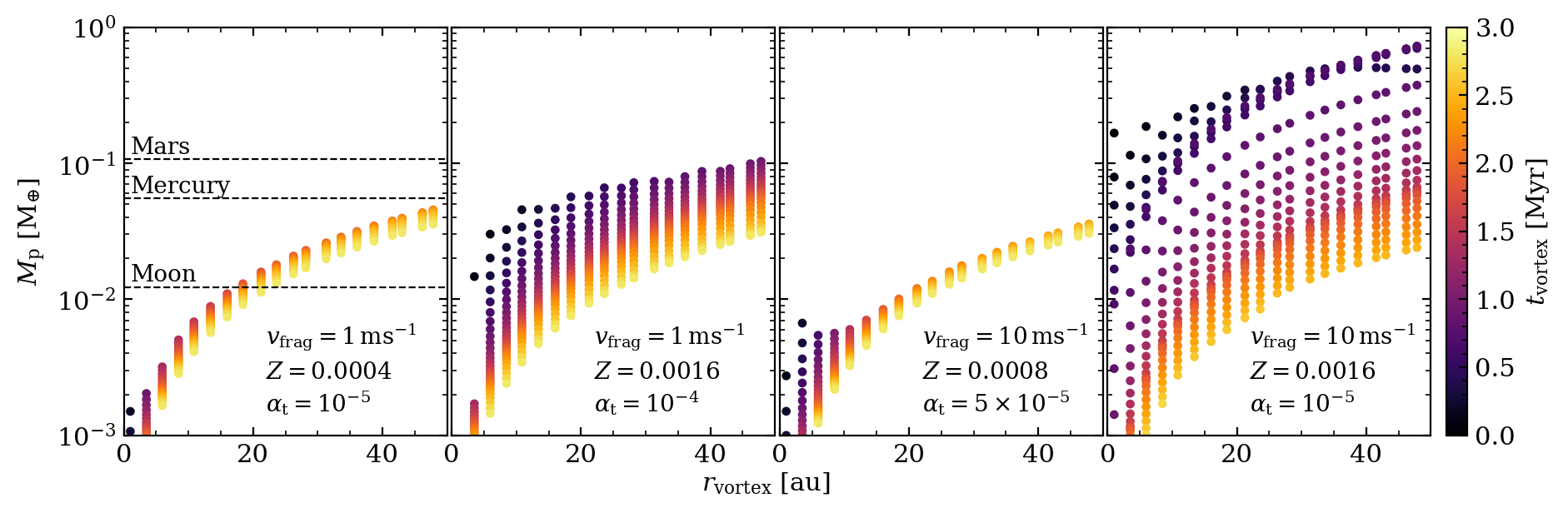}
    \caption{The plots display the final planetary masses as a function of the vortex formation location for a selection of the simulations shown in Fig. \ref{fig: mp}. The color of the scatter points represents the formation time of the vortex. }
    \label{fig: mp vs a}
\end{figure*}

Fig. \ref{fig: mp vs a} shows the masses of planets that are able to form as a function of the vortex formation location, for four different simulations.
The planetary masses increase with the vortex's formation distance from the star due to the strong dependence of trapped dust mass on semimajor axis ($M_{\rm trapped} \propto H^3$). If growth were dominated by standard pebble accretion outside the vortex, this trend would be reversed, as the pebble accretion rate decreases with semimajor axis. On average, planetary masses decrease with increasing vortex formation time. This is primarily because the background dust density peaks early in the disk’s evolution and decreases over time, resulting in vortices that form later having a smaller dust budget. Fig. \ref{fig: mp vs a} highlights that the timing and location of vortex formation within the disk significantly impact the masses of the planets forming inside it. 

To summarize, we find that early planet formation inside vortices is possible, provided that the forming embryo remains inside the vortex long enough to sweep up the trapped dust. If the embryo leaves the vortex before this occurs, little to no growth via pebble accretion takes place, and the embryo remains small. To form Mercury-mass planets or larger, $Z\gtrsim 0.04\, Z_{\odot}$ and a low level of turbulence are typically required. These results were obtained for a system with a stellar mass of $0.45\, \textrm{M}_{\odot}$ and a disk-to-star mass ratio of 1.6\%.
In Appendix \ref{sec:appendix:1Msun and varying disk-to-star mass}, we present the results of simulations performed with a stellar mass of $1\, \textrm{M}_{\odot}$ and disk-to-star mass ratios of 1\% and 10\%. When considering a disk-to-star mass ratio of 1\%, the results do not change significantly. However, when the disk-to-star mass ratio is increased to 10\%, planets with masses larger than Earth form for $Z\geq 0.08Z_{\odot}$ and low levels of turbulence.

\section{Discussion}\label{sec: discussion}
\subsection{Comparison with previous work}
In this study, we investigate whether vortices can trigger the formation of the first planets and planetesimals, and at what metallicity this could have occurred. Although the exact metallicity threshold varies with the assumed model and model parameters, most of our simulations suggest that planetesimals and Mercury-mass planets can form inside vortices when $Z\gtrsim 0.04Z_{\odot}$, provided a low level of turbulence. Previous studies on the critical metallicity threshold for planet formation include \citealt{Johansen2009}, \citet{Johnson2012}, \citet{Carrera2015} and \citet{Andama2024}.

\citet{Johnson2012} compute the timescales for dust growth and settling as a function of metallicity and semimajor axis, and compare these timescales to that for disk photoevaporation. By arguing that dust growth and settling must occur within the lifetime of the protoplanetary disk in order for planetesimals to form, they derive the following critical metallicity threshold:
\begin{equation}
    [\rm{Fe}/\rm{H}]_{\rm crit} \simeq - 1.5 + \log(r/ 1\, \rm{au}).
\end{equation}
This formula predicts a much higher metallicity threshold at large semimajor axes than what is found in our study. However, we note that this formula results in $[\rm{Fe}/\rm{H}]_{\rm crit}>0$ at $r>30\, \textrm{au}$, which is inconsistent with the existence of the cold classical Kuiper Belt, which is thought to have formed in-situ \citep{ParkerKavelaars10,Batygin+11}. Furthermore, distant debris disks have been detected around multiple stars, and it seems unlikely for all of those to have formed via the scattering of planetesimals from the inner disk.

\citet{Andama2024} simulate the evolution of viscous accretion disks while accounting for dust coagulation and the evaporation and condensation of chemical species at ice lines. They find that planetesimals form around the water ice line and identify a metallicity threshold at $[\textrm{Fe/H}]=-0.6$, corresponding to $Z\sim 0.25Z_{\odot}$. This threshold agrees well with the host star metallicity cutoff at $‑0.75 < [\textrm{Fe/H}] \leq ‑0.5$ identified by \citet{Boley2024} for the occurrence rate of super-Earths. In our model, super-Earths only form in systems with a disk-to-star mass ratio of 10\% and  $Z\gtrsim 0.1Z_{\odot}$. When the disk-to-star mass ratio and/or metallicity are lower, the planets that form are of lower mass than super-Earths. Furthermore, since planetary masses increase with semimajor axis, and common exoplanet detection methods are most sensitive for planets on close-in orbits, the planets formed in our study would be very hard to detect.

In a recent study by \citet{Whalen2025}, they model the formation of a dense cloud core formed in the debris of a pair-instability SNe explosion and the subsequent collapse of this cloud core to a protoplanetary disk. This disk forms at $z\sim 17$ and has a metallicity $Z=0.04\, Z_{\odot}$. The disk is subject to gravitational instability (GI), and a dead zone is formed in the inner few au due to the radially increasing strength of the GI. The trapping of dust inside the dead zone raises the dust-to-gas ratio above the SI criteria, triggering the formation of a ring of planetesimals. The total mass of planetesimals in this ring is sufficient to form Earth-mass planets via planetesimal collisions. 
Although the mechanism for dust trapping, numerical modeling, and system properties vary greatly between our study and that of \citet{Whalen2025}, both studies find planetesimal formation in a system with $Z=0.04\, Z_{\odot}$. 

\subsection{Limitations of the model}
Our study naturally has assumptions and simplifications. In the following discussion, we mention some of the main uncertainties of this study, and how they might impact the results. 

\subsubsection{Mono-disperse VS poly-disperse size distribution}\label{subsec: mono vs poly}
Although the unperturbed disk evolution and pebble accretion were performed with the full dust size distribution, we used a representative $\St$ when calculating the amount of growth and concentration inside a vortex, and when deciding whether or not planetesimals form. \citet{Li2020} find that efficient dust coagulation, fragmentation and drift towards the vortex center produces a wide range of grain sizes inside the vortex. Towards the vortex center, their maximum grain size is significantly larger than in other parts of the vortex, and the density-size distribution also shows a clear excess at the largest sizes (see their Fig. 2). Since the largest grains contain a significant fraction of the total dust mass, our approach to represent the entire size distribution by the density-weighted average $\St$ is reasonable, but even so the results would be affected by considering the entire size distribution. 

Since the peak dust-to-gas density ratio at the vortex center depends positively on $\St$ (see Eq. \ref{eq: vortex-boost}), taking the presence of smaller grains into account would somewhat lower the obtained ratio. In Fig. 2 of \citet{Li2020}, it can be seen that the dust-to-gas ratio at the vortex center increases when switching from using the poly-disperse size distribution to using the density-weighted average $\St$. Therefore, the planetesimal formation regions found in our study would likely become narrower if the entire size distribution was considered. 

Regarding the SI criteria used in this work, both are derived from simulations using a mono-disperse size distribution. Linear studies of the SI with a poly-disperse size distribution show that the SI is sometimes suppressed \citep{Krapp2019}; however, \citet{ZhuYang2021} and \citet{YangZhu2021} found that it can still operate at high dust-to-gas ratios or $\St$. More recently, \citet{Ho2024} presented simulations of the polydisperse SI with dust coagulation taken into account. They find that the interplay between dust coagulation and the SI can result in planetesimal formation even for relatively small initial dust sizes. Given the large grain sizes and dust-to-gas ratios expected at the vortex center, it is reasonable to assume that the SI can still operate. How the SI criteria change from the mono-disperse to the poly-disperse case remains an active area of research.

\subsubsection{Lifetime of the vortex}\label{subsec: lifetime vortex}
In this study, we assume a vortex lifetime of 1000 orbital periods when calculating the peak dust-to-gas ratios and Stokes numbers in case VCG. Increasing the vortex lifetime would result in additional growth and concentration, allowing planetesimals to form over a wider region than in our main study. Conversely, shortening the vortex lifetime would lead to narrower planetesimal formation regions. In Fig. \ref{fig:appendix:embryo tvortex}, we show how the planetesimal formation regions change when a 10 times longer vortex lifetime is considered. 

Our model does not account for the drift of grains into the vortex, or the change in gas density over time. The additional grains and lower gas densities would lead to higher dust-to-gas ratios, resulting in more planetesimal formation. Therefore, even if the vortex lifetime was shorter than assumed in our model, taking these effects into account might still result in similar or even larger planetesimal formation regions. 

\subsubsection{Mass of the formed embryo}
We made the assumption that the forming embryos are 10 times more massive than the typical planetesimals formed via the SI. In the simulations of \citet{Lyra2024}, embryos form with a wide range of masses, with the largest objects having masses similar to that of Mars. Since we assume a residence time of 10 orbital periods within the vortex, during which all trapped grains are accreted onto the embryo, increasing or decreasing the embryo mass would not affect our results. If the embryo leaves the vortex immediately after formation, as in the upper panel of Fig. \ref{fig: mp_single}, increasing the embryo mass would result in higher pebble accretion rates and larger final planetary masses. Due to the low solid density of the disk, the total accreted mass would likely still be low though. 

\subsubsection{The formation of multiple embryos}\label{subsubsec: multiple embryos}
In this study, we assumed that a single embryo formed inside each vortex, and this embryo then proceeds to accrete the remaining dust mass within the vortex. Since the criteria for forming planetesimals is only reached towards the vortex center in our low-metallicity disks, if multiple planetesimals form they likely do so relatively close to each other. If they were to collide and participate in the formation of a single embryo, the results would not change compared to our model. If multiple planetesimals/embryos form and leave the vortex before they have a change to collide, the remaining dust mass trapped within the vortex would be lower than assumed in our model, and thus the final planetary masses would be lower.

\section{Conclusion}\label{sec: conclusion}
We have investigated whether the presence of vortices can trigger the formation of the first planets and planetesimals in the early universe. Our model incorporates dust evolution, enhancement of dust growth and concentration inside vortices, planetesimal formation via the SI, and pebble accretion. We conducted simulations across a range of metallicities and turbulence levels to identify the critical metallicity threshold.

We identify a metallicity threshold of $Z\gtrsim 0.04\, Z_{\odot}$ for planetesimal formation, which could potentially extend to even lower metallicities in the case of long-lived vortices. Within the range of parameters explored in this study, this threshold remains unchanged regardless of the assumed fragmentation velocity, SI criteria, stellar mass, or disk-to-star mass ratio. However, the threshold is heavily dependent on the turbulence level. The quoted threshold value was obtained for $\alpha_{\rm turb}=10^{-5}$, with higher turbulence levels leading to higher metallicity thresholds. 
 
If the formed planetesimals/embryos remain inside the vortex long enough to accrete all of the trapped dust, Mercury-mass planets ($0.055\, \textrm{M}_{\oplus}$) can form even at a metallicity of $Z\sim0.04\, Z_{\odot}$. To form Mars-mass planets ($0.107\, \textrm{M}_{\oplus}$) or larger, a metallicity of $Z\gtrsim 0.08\, Z_{\odot}$ is required. Conversely, if the embryo leaves the vortex immediately after formation, mass growth via pebble accretion is negligible, and the embryos remain small.

Studies have shown that low-mass, metal-enriched stars can form already at cosmic dawn. \citet{Whalen2025} find that a protoplanetary disk with $Z=0.04\, Z_{\odot}$ can form in the debris of a single pair-instability supernova. Since our results show planetesimal formation occurring in disks with $Z=0.04\, Z_{\odot}$, this suggests that vortices could trigger the formation of the first generation of planets and planetesimals in the universe.

In the future, a study constraining the timescales for dust growth and concentration inside vortices would be necessary to better refine the metallicity threshold and its dependence on turbulence. Furthermore, our prescriptions for planetesimal formation and dust concentration inside vortices are based on simulations with single dust sizes. Deriving criteria that account for a polydisperse dust size distribution would provide more accurate metallicity thresholds. 

\section*{Acknowledgements}
The authors wish to thank the anonymous referee for providing helpful comments that lead to an improved manuscript. 
LE acknowledges the support from NASA via the Emerging Worlds program (\#80NSSC25K7117), as well as funding by the Institute for Advanced Computational Science Postdoctoral Fellowship. WL acknowledges support from the NASA Theoretical and Computational Astrophysical Networks (TCAN) via grant \#80NSSC21K0497. WL is further supported by grants \#80NSSC22K1419 from the NASA Emerging Worlds program and AST-2007422 from NSF. DC and WL acknowledge support from NASA under {\em Emerging Worlds} through grant 80NSSC25K7414 and {\em Exoplanets Research Program} through grant 80NSSC24K0959. B.B. acknowledges support from NSF grant AST-2009679 and 2407877.  B.B. and S.M. acknowledge support from NASA grant No. 80NSSC20K0500.
B.B. is grateful for generous support from the David and Lucile Packard Foundation and the Alfred P. Sloan Foundation.

\section*{Data Availability}
All simulations and results can be reproduced using our open-source code firstPlanets, available on GitHub: https://github.com/astrolinn/firstPlanets. Furthermore, the data underlying this article will be shared on reasonable request to the corresponding author.



\bibliographystyle{mnras}
\bibliography{refs}




\appendix 


\section{Calculation of residence time inside the vortex}
\label{sec:appendix:timeInVortex}

A planetesimal forming within the vortex is large enough to be mostly unaffected by gas drag, and thus orbits at the Keplerian velocity. If the planetesimal forms at exactly the same semimajor axis as the vortex center, it will have the same orbital velocity as the vortex and should in theory never leave the vortex. Any radial displacement from the vortex center will result in a final residence time within the vortex, as the difference in orbital velocity will result in the planetesimal eventually drifting out of the vortex. 

The vortex can be described by an ellipse with semi-minor axis $a$ and semimajor axis $b=a\chi$, where $\chi=4$ and $a=2H/3$ for the Kida solution. Let the vortex center have coordinates $(x,y)=(0,0)$, where x points in the direction of the semi-minor axis (the radial direction) and y points in the direction of the semimajor axis (the direction along the orbit).
The equation for an ellipse is:
\begin{equation}
    \frac{(\Delta x)^2}{a^2} + \frac{(\Delta y)^2}{b^2} = 1,
\end{equation}
where $\Delta x$ is the radial displacement from the vortex center, and $\Delta y$ is the corresponding distance from $(\Delta x,0)$ to the edge of the vortex. The maximum distance a planetesimal forming at a distance $\Delta x$ from the vortex center could travel before it leaves the vortex is:
\begin{equation}
    2\Delta y = 2b\sqrt{1-\frac{(\Delta x)^2}{a^2}}.
\end{equation}

From the shearing sheet approximation, we get that the difference in orbital velocity between two points with radial separation $\Delta x$ is:
\begin{equation}
    \Delta v = \frac{3}{2}\Omega_{\rm K}\Delta x = \frac{3\pi \Delta x}{P},
\end{equation}
where $P$ is the orbital period at the vortex center. The maximum crossing time for a planetesimal forming at radial distance $\Delta x$ from the vortex center can then be calculated as:
\begin{equation}
    t_{\rm cross} = \frac{2\Delta y}{|\Delta v|} = \frac{2\Delta y}{3\pi \Delta x}P.
\end{equation}

\begin{figure}
    \centering
    \includegraphics[width=\columnwidth]{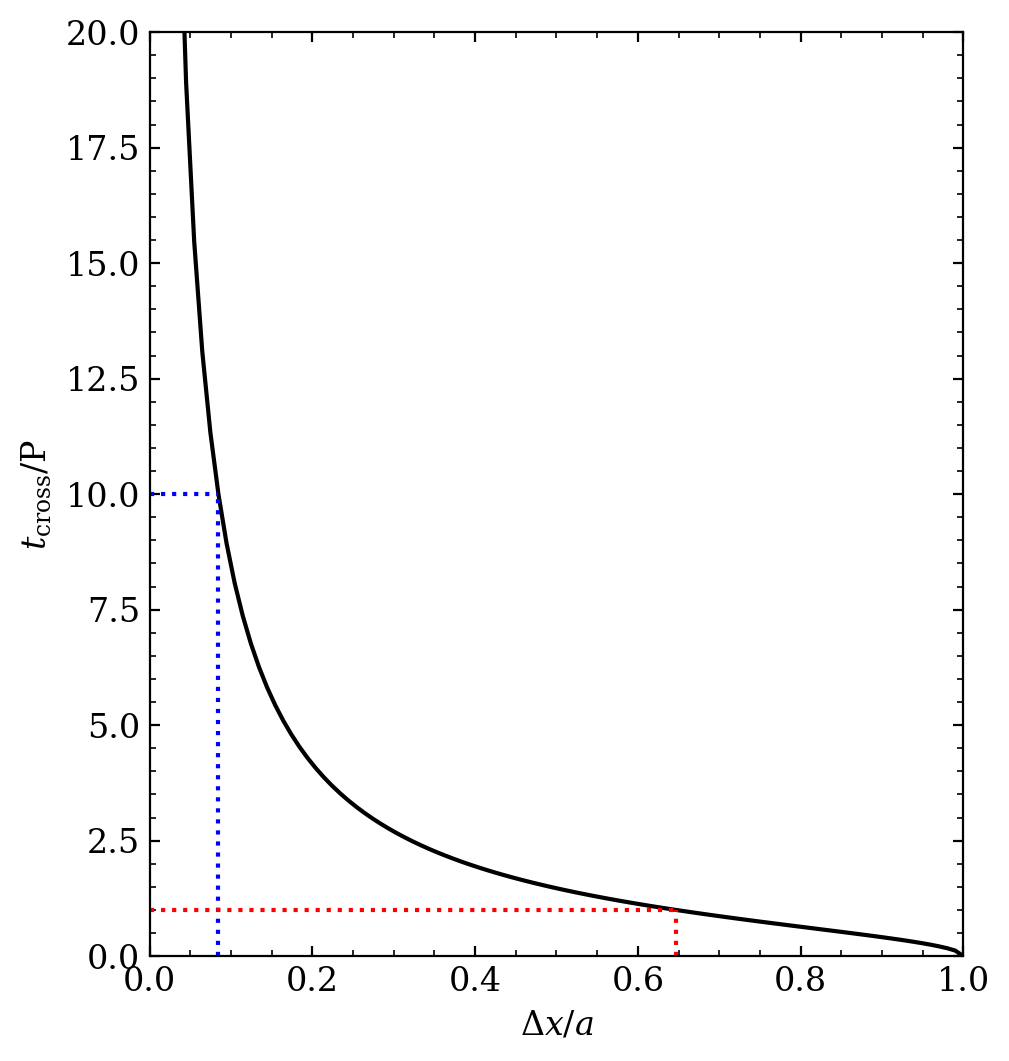}
    \caption{Maximum crossing time of an embryo forming inside the vortex as a function of radial separation from the center of the vortex.}
    \label{fig: tcross}
\end{figure}

In Fig. \ref{fig: tcross} we show the crossing time as a function of radial separation from the vortex center. In our main study we assume that the planetesimals reside 10 orbital periods within the vortex, which corresponds to a maximum radial separation of $\Delta x/a \sim 0.08$ from the vortex center. 

\section{Increasing the vortex lifetime}\label{appendix:sec:tvortex}

\begin{figure*}
    \centering
    \includegraphics[width=1\columnwidth, trim=0.2cm 0.2cm 2.8cm 0, clip]{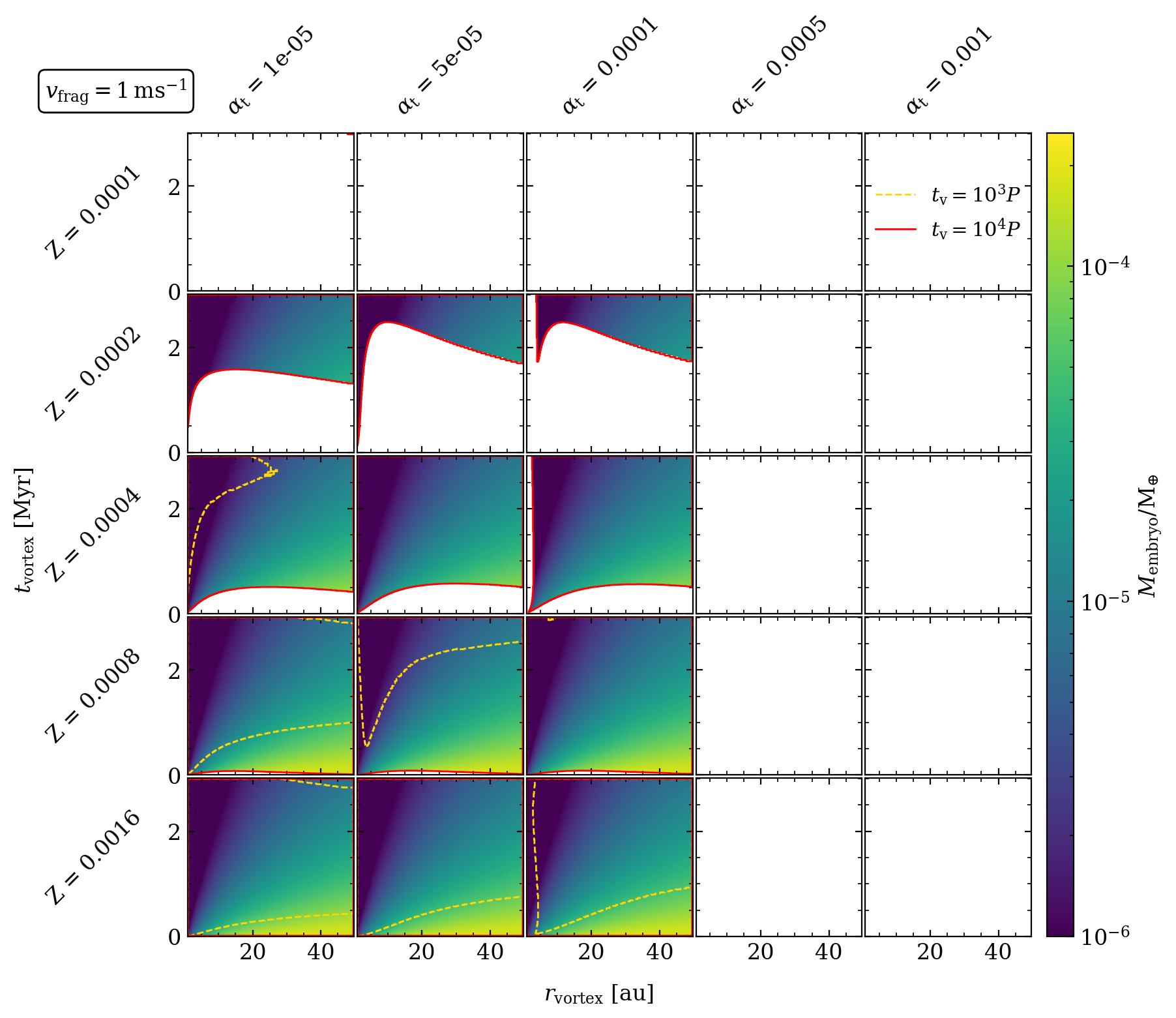}
    \includegraphics[width=1\columnwidth, trim=2.8cm 0.2cm 0.2cm 0, clip]{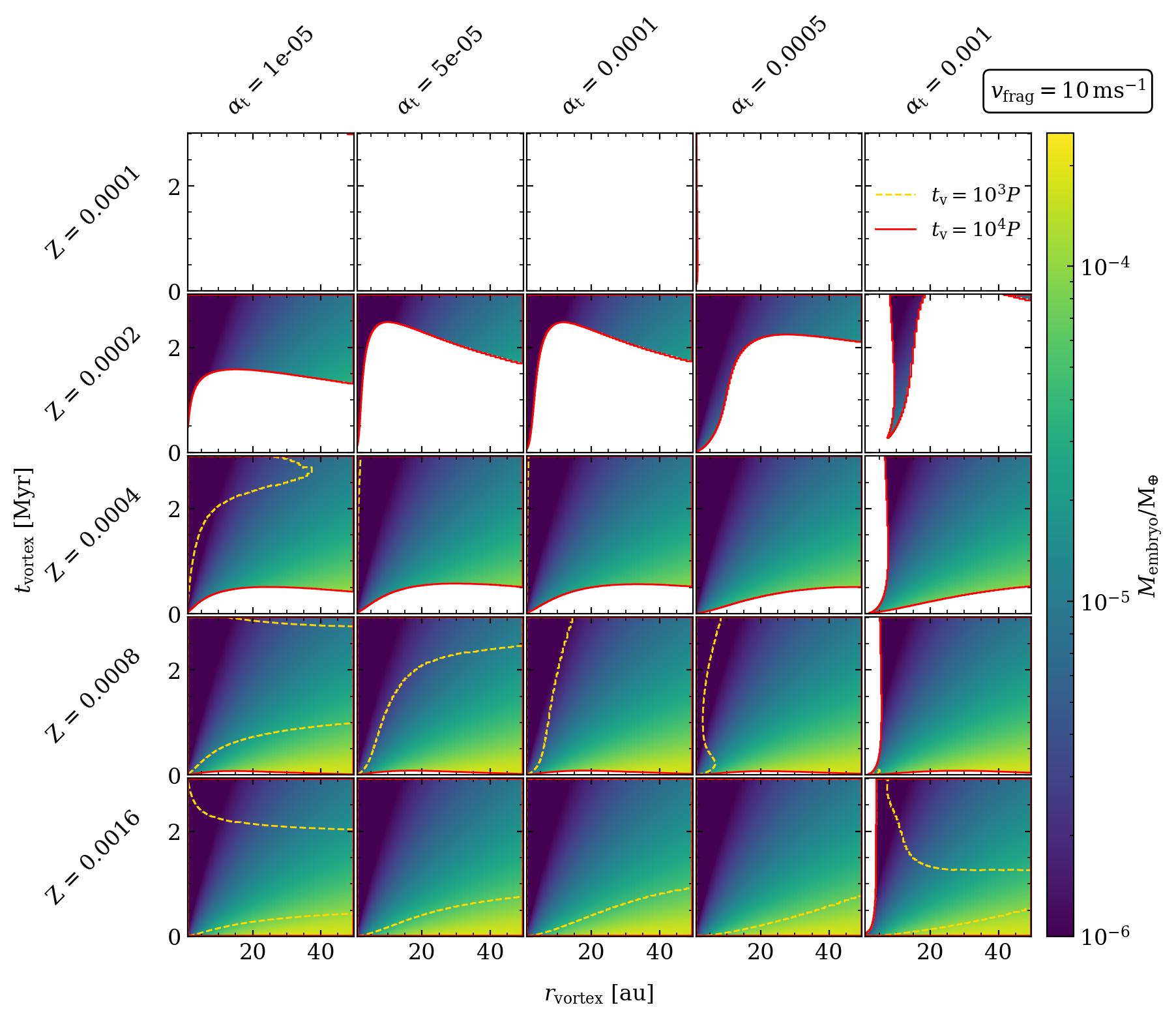}
    \caption{The plots show how the region where vortex formation results in planetesimal formation changes when different vortex lifetimes are assumed. These results are obtained using SI criteria SI24. }
    \label{fig:appendix:embryo tvortex}
\end{figure*}

In our main study, we assumed a vortex lifetime of 1000 orbital periods when calculating the peak dust-to-gas ratio and Stokes number obtained at the vortex center in case VCG. We performed additional simulations with a 10 times longer vortex lifetime, the results of which are shown in Fig. \ref{fig:appendix:embryo tvortex}. In this scenario, planetesimal formation becomes possible also for $Z= 0.02Z_{\odot}$. When comparing simulations with $v_{\rm frag}=1\, \textrm{ms}^{-1}$ and $v_{\rm frag}=10\, \textrm{ms}^{-1}$, there is little difference when the turbulence level is low. For high levels of turbulence, the SI criteria can be met when $v_{\rm frag}=10\, \textrm{ms}^{-1}$, but not when $v_{\rm frag}=1\, \textrm{ms}^{-1}$. 


\section{Simulations with a solar mass star and varying disk-to-star mass ratios}
\label{sec:appendix:1Msun and varying disk-to-star mass}

\begin{figure*}
    \centering
    \includegraphics[width=1\columnwidth, trim=0.2cm 0.2cm 2.8cm 0, clip]{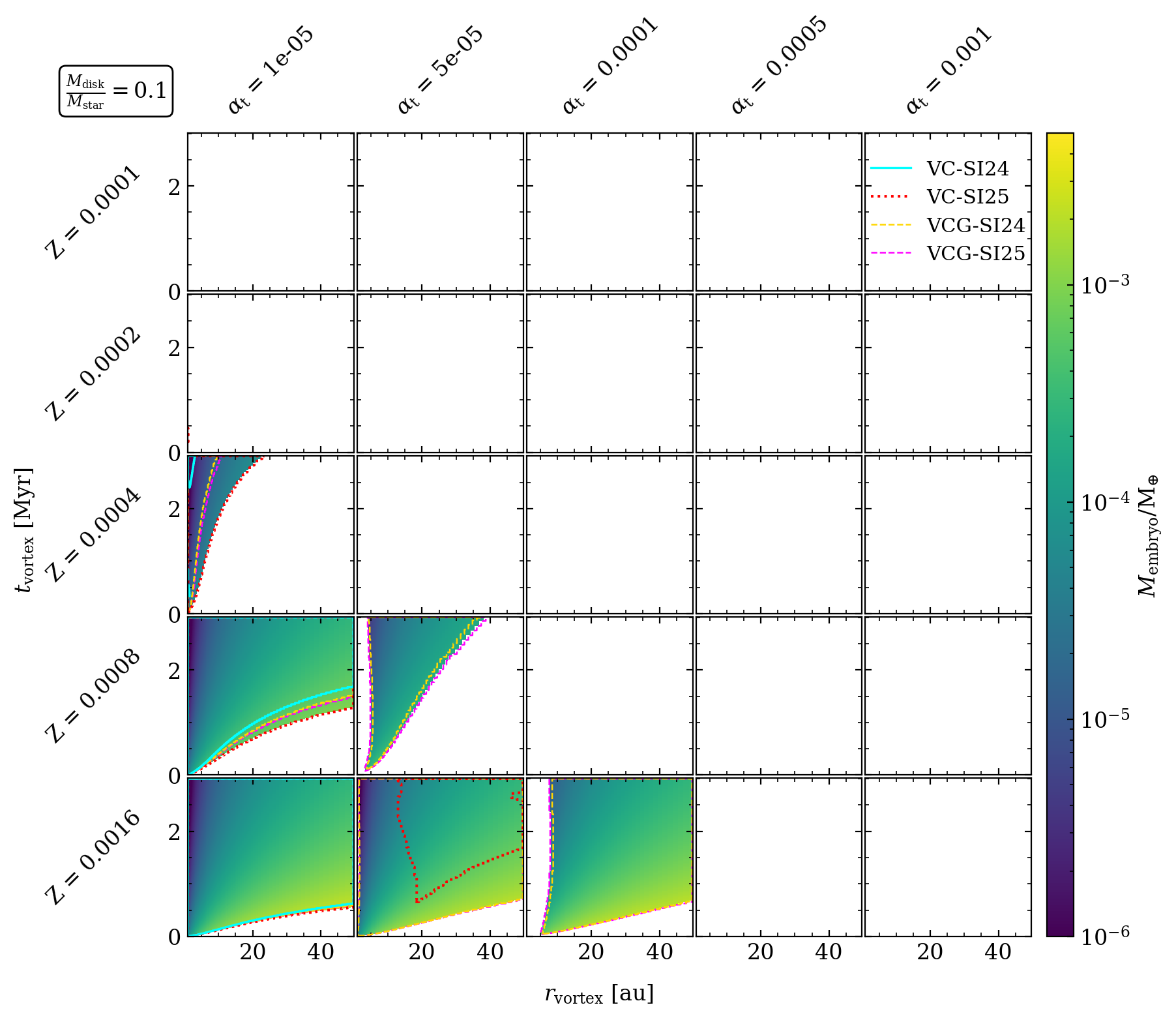}
    \includegraphics[width=1\columnwidth, trim=2.8cm 0.2cm 0.2cm 0, clip]{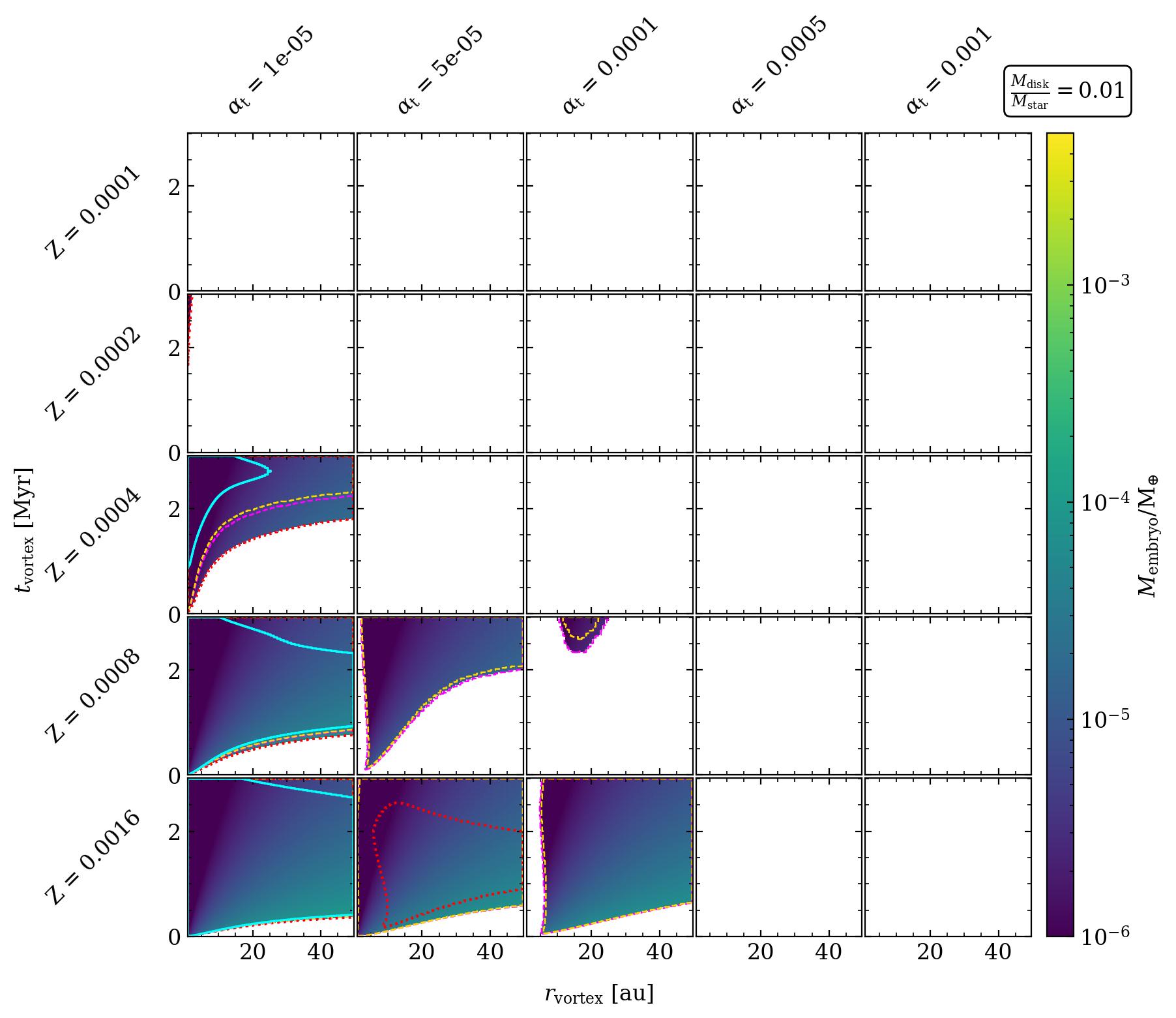}
    \caption{The plots show where on the space-time grid vortex formation results in planetesimal formation, when a stellar mass of $1\, \textrm{M}_{\odot}$ and a disk-to-star mass ratio of 10\% (left plot) and 1\% (right plot) are adopted.}
    \label{fig:appendix:membryo}
\end{figure*}

\begin{figure*}
    \centering
    \includegraphics[width=1\columnwidth, trim=0.2cm 0.2cm 2.8cm 0, clip]{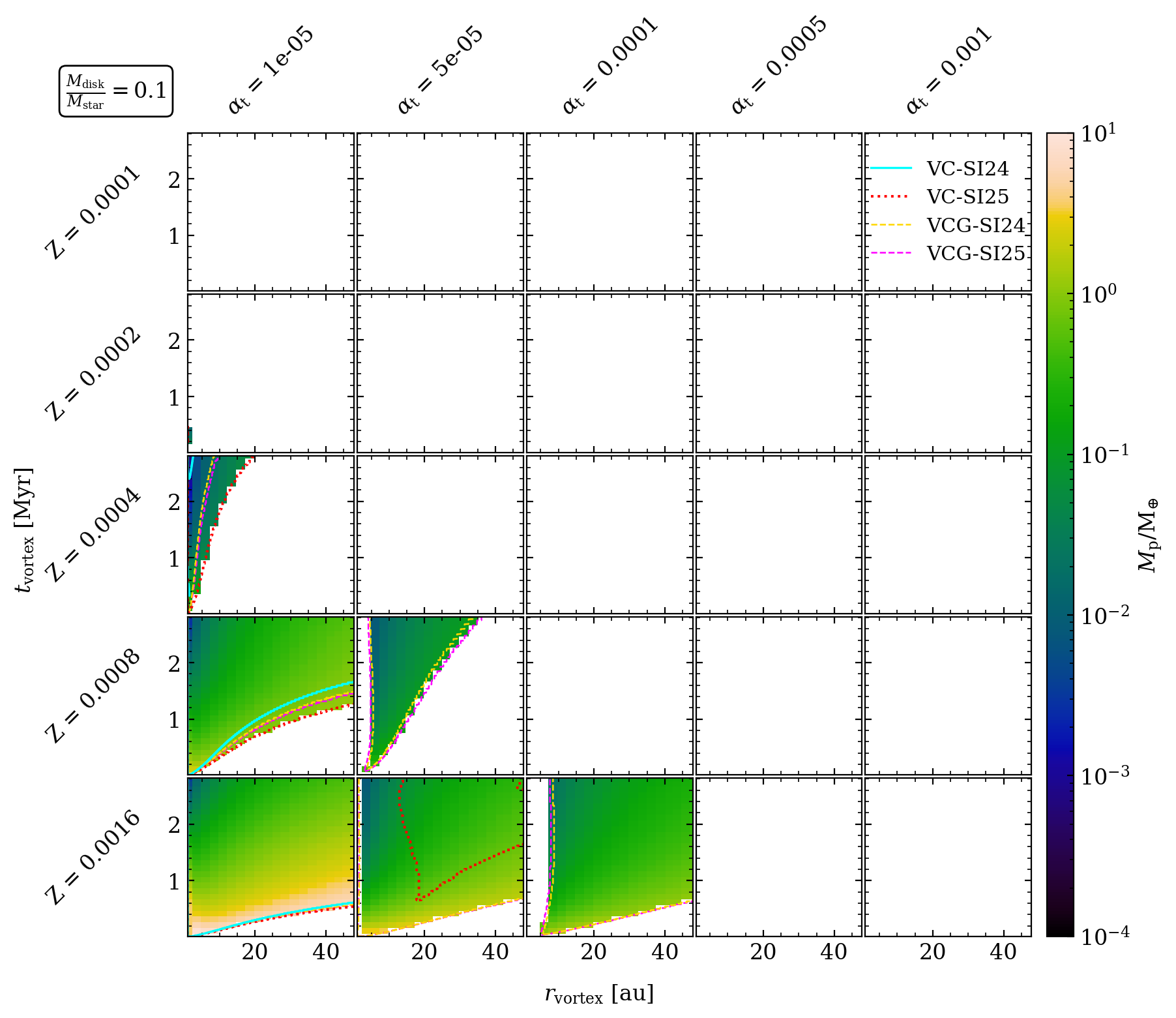}
    \includegraphics[width=1\columnwidth, trim=2.8cm 0.2cm 0.2cm 0, clip]{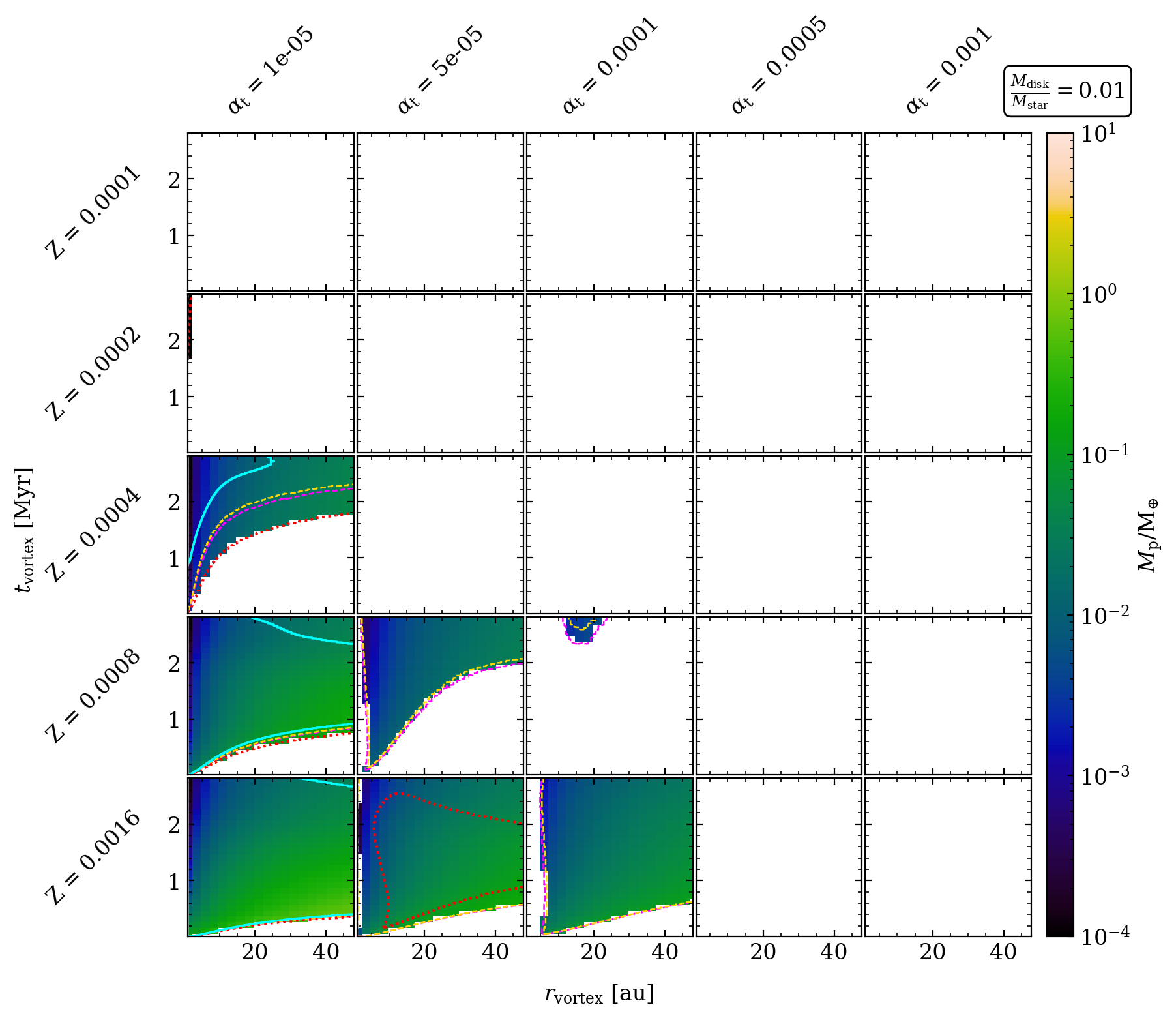}
    \caption{The plots show the final planetary masses for the embryos formed in Fig \ref{fig:appendix:membryo}, when assuming a residence time of 10 orbital periods within the vortex.}
    \label{fig:appendix:mp}
\end{figure*}

The main study was performed using a single stellar mass and disk-to-star mass ratio, which were obtained from an early star formation simulation. In the following, we present results from additional simulations performed with a solar mass star and disk-to-star mass ratios of 1\% and 10\%. The corresponding stellar radius and temperature are $R_*=3.096\, \textrm{R}_{\odot}$ and $T_*=4397\, \textrm{K}$, and the disk accretion rate is $1.65\times 10^{-8}\, \textrm{M}_{\odot}\textrm{yr}^{-1}$/$1.65\times 10^{-7}\, \textrm{M}_{\odot}\textrm{yr}^{-1}$ for a 1/10\% disk-to-star mass ratio, respectively. We adopt a fragmentation velocity of $1\, \textrm{ms}^{-1}$ in these simulations.

The locations on the space-time grid where planetesimal formation inside vortices is possible are shown in Fig. \ref{fig:appendix:membryo}. The regions of planetesimal formation do not vary much with the assumed disk-to-star mass ratio, and remain much the same as in the simulations in the main study. The masses of the formed embryos is significantly higher when the disk-to-star mass ratio is 10\%. 

The final planetary masses that are obtained when assuming a residence time of 10 orbital periods within the vortex are shown in Fig. \ref{fig:appendix:mp}. When considering a disk-to-star mass ratio of 1\%, the results remain much the same as in the main study. When the disk-to-star mass ratio is increased to 10\%, planets with masses above an Earth mass can form when $Z\geq0.08Z_{\odot}$ and the turbulence level is low. For the most massive planets that form, our assumption to ignore gas accretion and planetary migration is no longer valid. 

\section{Dust growth in vortices}
\label{sec:appendix:dust_growth}

We adopt the model of \citet{Carrera2025}, who found a powerful feedback loop between vortex concentration and dust coagulation. We refer to that paper for the details of the model. Here we simply summarize the key ideas and describe the final algorithm. Vortices have several properties that make them prime candidates for planet formation:

\begin{itemize}
\item Unbounded $Z$: Vortices are particle traps that concentrate dust grains toward their center. The dust-to-gas ratio $Z$ inside the vortex is, in principle, unbounded, as the steady flow of dust drifting through the protoplanetary disk is continuously trapped by the vortex.

\item No drift barrier: In general, planetesimal formation is thought to be impeded by two barriers, the radial drift barrier (outer disk) and the fragmentation barrier (inner disk). Because vortices are particle traps, they eliminate the drift barrier, leaving only fragmentation.

\item Turbulence dampening: Vortices are already known to have lower turbulence than their surroundings, independent of dust mass loading \citep{Lesur_2010,Lyra_2011}. The key insight of \citet{Carrera2025}) is that the amount of turbulence dampening is a dynamic quantity, that responds to grain growth occurring within the vortex.
\end{itemize}

\noindent
To keep the model both simple and conservative, we ignore the first bullet point and assume that the vortex's entire dust budget is fixed, and equal to only the dust present at its radial location at the time of formation. The model ingredients are simple enough:

\begin{itemize}
\item Turbulence dampening: \citet{Carrera2025} derived the turbulent gas velocity, $\Vg$, in a gas-dust mixture where both $\St$ and the midplane dust-to-gas ratio $\epsilon$ are non-negligible:

\begin{eqnarray}
    \Vg &=& \sqrt{\alpha}\ceff
    \label{eqn:Vgas}
    \\
    \ceff &\equiv& \cs \sqrt{\frac{1 + \St}{1 + \St + \epsilon}}
    \label{eqn:ceff}
\end{eqnarray}

\noindent
and they refer to $\ceff$ as the ``effective'' sound speed.

\item Fragmentation barrier: In this setting, the fragmentation barrier becomes

\begin{equation}\label{eqn:Stfrag}
    \St_{\rm frag}
    = \frac{V_{\rm frag}^2}{3\alpha\cs^2} \left(1 + \frac{\epsilon}{1 + \St} \right)
\end{equation}

\noindent
where $V_{\rm frag}$ is the fragmentation speed of the dust material. In the limit as $\epsilon \rightarrow 0$, this formula gives the familiar expression for the fragmentation barrier.

\item Vortex concentration: The vortex concentrates dust into a Gaussian density profile that peaks in the center of the vortex with column dust-to-gas ratio \citep{Lyra2013}

\begin{equation}\label{eqn:Zmax}
    Z_{\rm max} = Z_{\rm disk} \left(1 + \frac{\St}{\alpha}\right)
\end{equation}

\item Dust sedimentation: Lastly, mass loading also enhances dust sedimentation. This was seen already in the numerical simulations of \citet{Lim2024}. Again, we use the expression derived by \citet{Carrera2025}:

\begin{eqnarray}\label{eqn:epsilon}
    \epsilon &=& \frac{\zeta + \sqrt{\zeta^2 + 4\zeta(1+\St)}}{2}\\
    \zeta &\equiv& Z^2 (1+\St/\alpha) / (1+\St)
\end{eqnarray}

\end{itemize}

\noindent
In other words, higher $\epsilon$ leads to larger $\St_{\rm frag}$ (Equation \ref{eqn:Stfrag}), while larger $\St$ leads to higher $Z$ (Equation \ref{eqn:Zmax}) and $\epsilon$. To pair these processes, we need their characteristic timescales. \citep{Birnstiel_2012} derived the growth timescale $t_\St = 1/(Z\Omega)$. For a vortex we replace the Keplerian frequency with the vortex frequency $\Omegav \equiv 0.5\Omega$, giving

\begin{equation}\label{eqn:tSt}
    t_\St = \frac{1}{Z\,\Omegav} = \frac{2}{Z\,\Omega}
\end{equation}

\noindent
Lastly, \citet{Carrera2025} derived the vortex concentration timescale $t_Z = 1/(\St\Omega)$, but they noted that numerical simulations suggest a slightly higher $t_Z$. Therefore, we set

\begin{equation}\label{eqn:tZ}
     t_Z = \frac{1}{\St\,\Omegav} = \frac{2}{\St\,\Omega}
\end{equation}

\noindent
placing $t_Z$ more in line with numerical simulations. The obvious parallel in Equations \ref{eqn:tSt} and \ref{eqn:tZ} highlights the interdependence of these two processes. We write the final algorithm as Python code:

\begin{lstlisting}
# Equation D6
def epsilon(Z,St,alpha):
    x = Z**2 * (1+St/alpha) / (1+St)
    return (x + sqrt(x*x + 4*x*(1+St)))/2

# Main Loop
Z   = Z0  = Z_disk
St  = St0 = vfrag**2 / (3*alpha*Cs**2)
eps = epsilon(Z,St,alpha)
Omega_v = 0.5 # Vortex frequency

while (cond):
    Z_max   = Z0  * (1 + St/alpha)
    St_frag = St0 * (1+St+eps)/(1+St)
    
    # Timestep for this iteration
    t_Z  = 1/(St*Omega_v)
    t_St = 1/(Z *Omega_v)
    dt   = 0.1*min(t_Z, t_St)
    
    # Z and St both grow together
    Z   = min(Z_max  , Z  * exp(dt/t_Z ))
    St  = min(St_frag, St * exp(dt/t_St))
    eps = epsilon(Z,St,alpha)
\end{lstlisting}

\section{Streaming Instability Criterion}
\label{sec:appendix:dust_growth:si_criterion}

The SI criterion from \citet{Lim2025} (SI25) is expressed in terms of the dust-go-gas surface density:

\begin{equation}
    \log(Z_{\rm crit}) \approx 0.10 (\log \St)^2
    + 0.07 \log \St - 2.36.
\end{equation}

\noindent
To use this criterion, we must convert it into the equivalent midplane dust-to-gas density $\epsilon_{\rm crit}$. We can write the midplane density as $\epsilon \approx Z (H/\Hp)$, where $\Hp$ is the scale height of the dust layer. \citet{Lim2024} found that

\begin{equation}
    \Hp \approx \Heff \sqrt{
        \left(\frac{\Pi}{5}\right)^2 +
        \frac{\alpha}{\alpha + \St}
    }
\end{equation}

\noindent
where $\Pi$ is a parameter that quantifies the strength of the pressure gradient and $\Heff = H/\sqrt{1 + \epsilon}$ is the ``effective'' scale height of the gas-dust mixture. The $(\Pi/5)^2$ term is an estimate of the self-generated turbulence due to the SI. The simulations of \citet{Lim2025} had $\Pi = 0.05$ and no forced turbulence. Therefore, their simulations had 

\begin{equation}
    \Hp \approx \Heff \left(\frac{\Pi}{5}\right)
        = 0.01\Heff
\end{equation}

\noindent
At this point we deviate slightly from \citet{Lim2024}. Their expression for $\Heff$ was based on approximating the gas-dust mixture as a colloid ($\St \ll 1$), but we are interested in the case where $\St$ is not small. Therefore, we use the formula from \citet{Carrera2025}, who derived $\Heff$ without the colloid approximation,

\begin{equation}
    \Heff = H \sqrt{\frac{1 + \St}{1 + \St + \epsilon}}
\end{equation}

\noindent
Notice that, in the limit as $\St \rightarrow 0$, this expression becomes identical to the one in \citet{Lim2024}. Considering the small $\St$ values in their simulations, both expressions are equally consistent with their dataset. Altogether, the midplane dust-to-gas ratio is

\begin{equation}
    \epsilon =
        Z\frac{H}{\Hp}
        = 100 Z \sqrt{\frac{1 + \St + \epsilon}{1 + \St}}
\end{equation}

\noindent
Let $\xi \equiv 100Z_{\rm crit}/\sqrt{1+\St}$, and we get

\begin{equation}
    \epsilon_{\rm crit} = \frac{\xi^2 + \sqrt{\xi^4 + 4\xi^2(1+\St)}}{2}
\end{equation}


\bsp	
\label{lastpage}
\end{document}